\pdfoutput=1
\documentclass[journal,twopages]{IEEEtran}
\usepackage{epsfig}
\usepackage{epstopdf}
\usepackage{graphicx}
\usepackage{subfigure}
\usepackage{amsmath,amssymb}
\usepackage{color}
\usepackage{pslatex}
\usepackage{url}
\usepackage{psfrag}
\usepackage{cite}
\usepackage{algorithm}
\usepackage{url}
\usepackage[noend]{algorithmic}
\usepackage{adjustbox}
\usepackage{setspace}
\usepackage[font=small]{caption}
\usepackage{fancyhdr}
\usepackage{hyperref}

\long\def\symbolfootnote[#1]#2{\begingroup%
\def\thefootnote{\fnsymbol{footnote}}\footnote[#1]{#2}\endgroup}

\newif\ifarxiv
\arxivfalse

\newcommand{\makered}[1]{\color{black}#1\color{black}}

\newcommand{\comment}[1]{}

\newenvironment{myitemize}{\begin{list}{$\bullet$}{\renewcommand{\leftmargin}{0.2in}}}{\end{list}}

\begin{document}

%
%
\markboth{Margolies \MakeLowercase{\textit{et al.}}: Project-based Learning within a Large-Scale Interdisciplinary Research Effort}{}
\title{Project-based Learning within a Large-Scale Interdisciplinary Research Effort\vspace*{0.2in}}

\author{Robert~Margolies, 
        Maria~Gorlatova, 
        John~Sarik, 
        Peter~Kinget, 
        Ioannis~Kymissis, 
        and~Gil~Zussman 
\thanks{R.~Margolies, M.~Gorlatova, J.~Sarik, P.~Kinget, I.~Kymissis, and G.~Zussman are with the Department of Electrical Engineering, Columbia University, New York, NY 10027. E-mail: rsm2156@columbia.edu, maria.gorlatova@gmail.com, jsarik@gmail.com,
\{kinget, johnkym, gil\}@ee.columbia.edu.}%
\thanks{This work was supported in part by Vodafone Americas Foundation Wireless Innovation Project, United Microelectronics, a grant from Texas Instruments, and NSF grants CCF-0964497, CNS-0916263, CNS-10-54856.}
}


%
%
%


\maketitle


\begin{abstract}
The modern engineering landscape increasingly requires a range of skills to successfully integrate complex systems. \mbox{Project-based} learning is used to help students build professional skills. However, it is typically applied to small teams and small efforts. This paper describes an experience in engaging a large number of students in research projects within a \mbox{multi-year} interdisciplinary research effort. The projects expose the students to various disciplines in Computer Science (embedded systems, algorithm design, networking),
Electrical Engineering (circuit design, wireless communications, hardware prototyping), and Applied Physics (\mbox{thin-film} battery design, solar cell fabrication). While a student project is usually focused on one discipline area, it requires interaction with at least two other areas. Over 5 years, 180 \mbox{semester-long} projects have been completed. The students were a diverse group of high school, undergraduate, and M.S.\ Computer Science, Computer Engineering, and Electrical Engineering students.
Some of the approaches that were taken to facilitate student learning are real-world system development constraints, regular cross-group meetings, and extensive involvement of Ph.D.~students in student mentorship and knowledge transfer. To assess the approaches, a survey was  conducted among the participating students. The results demonstrate the effectiveness of the approaches. For example, 70\% of the students surveyed indicated that working on their research project improved their ability to function on multidisciplinary teams more than coursework, internships, or any other activity.
\end{abstract}


\IEEEpeerreviewmaketitle

\section{Introduction}
\begin{figure}[t]
\centering
\hspace{.2in}
\subfigure[\label{fig:mockup}]{\includegraphics[height = 0.3\columnwidth]{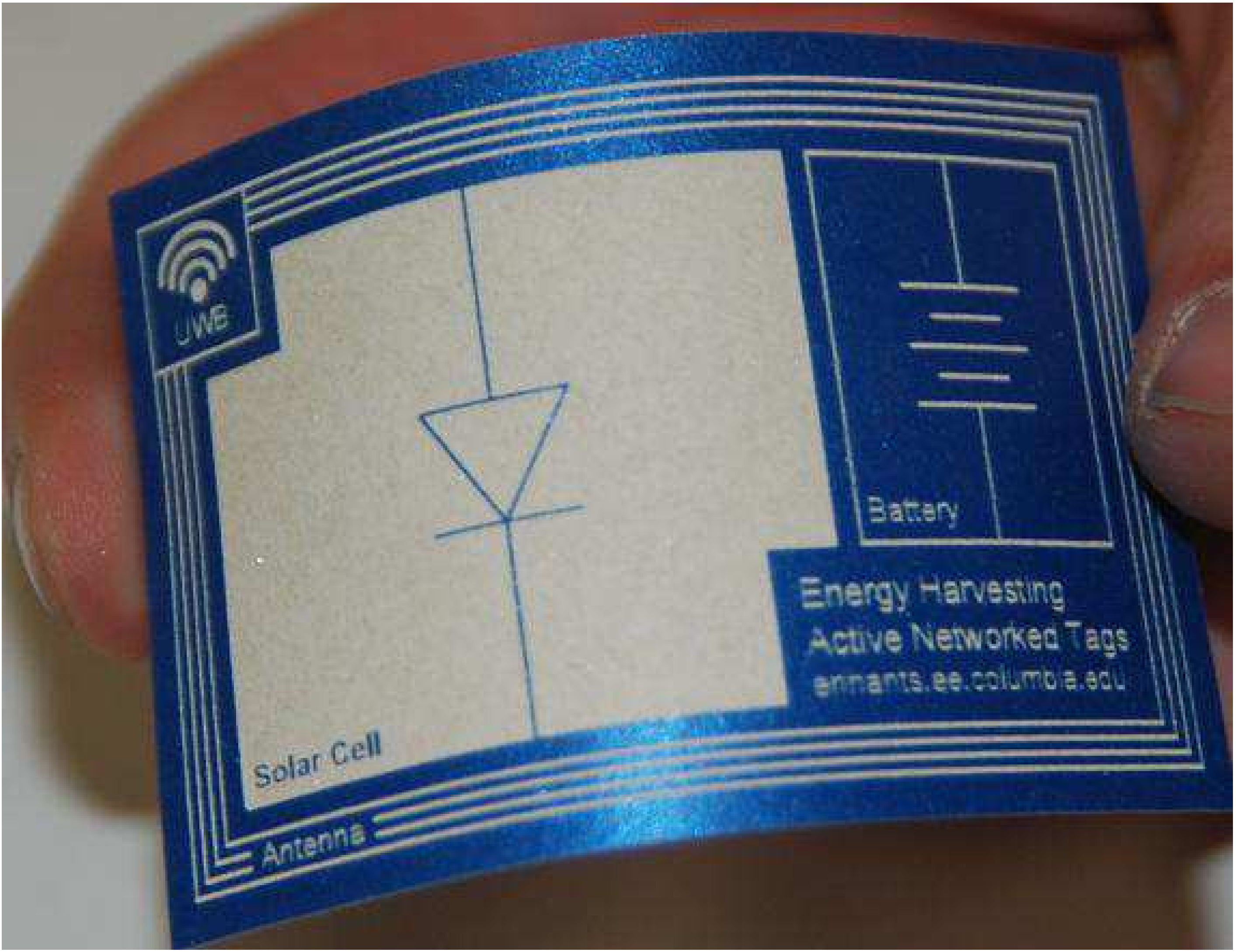}}
\hspace{.2in}
\subfigure[\label{fig:mckup2}]{\includegraphics[height = 0.3\columnwidth]{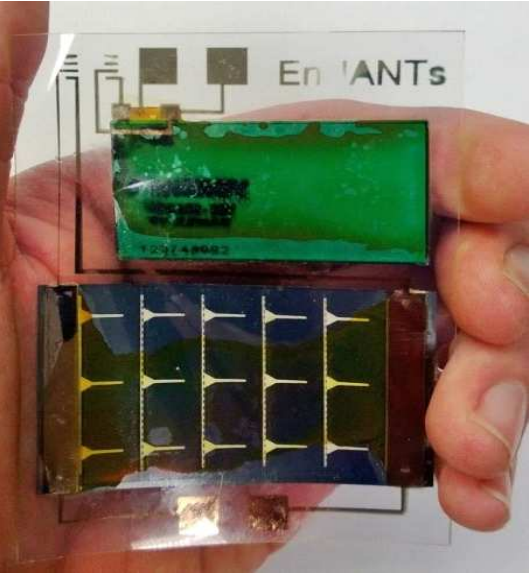}}
\hspace{.2in}
\subfigure[\label{fig:network}]{\includegraphics[height=0.3\columnwidth]{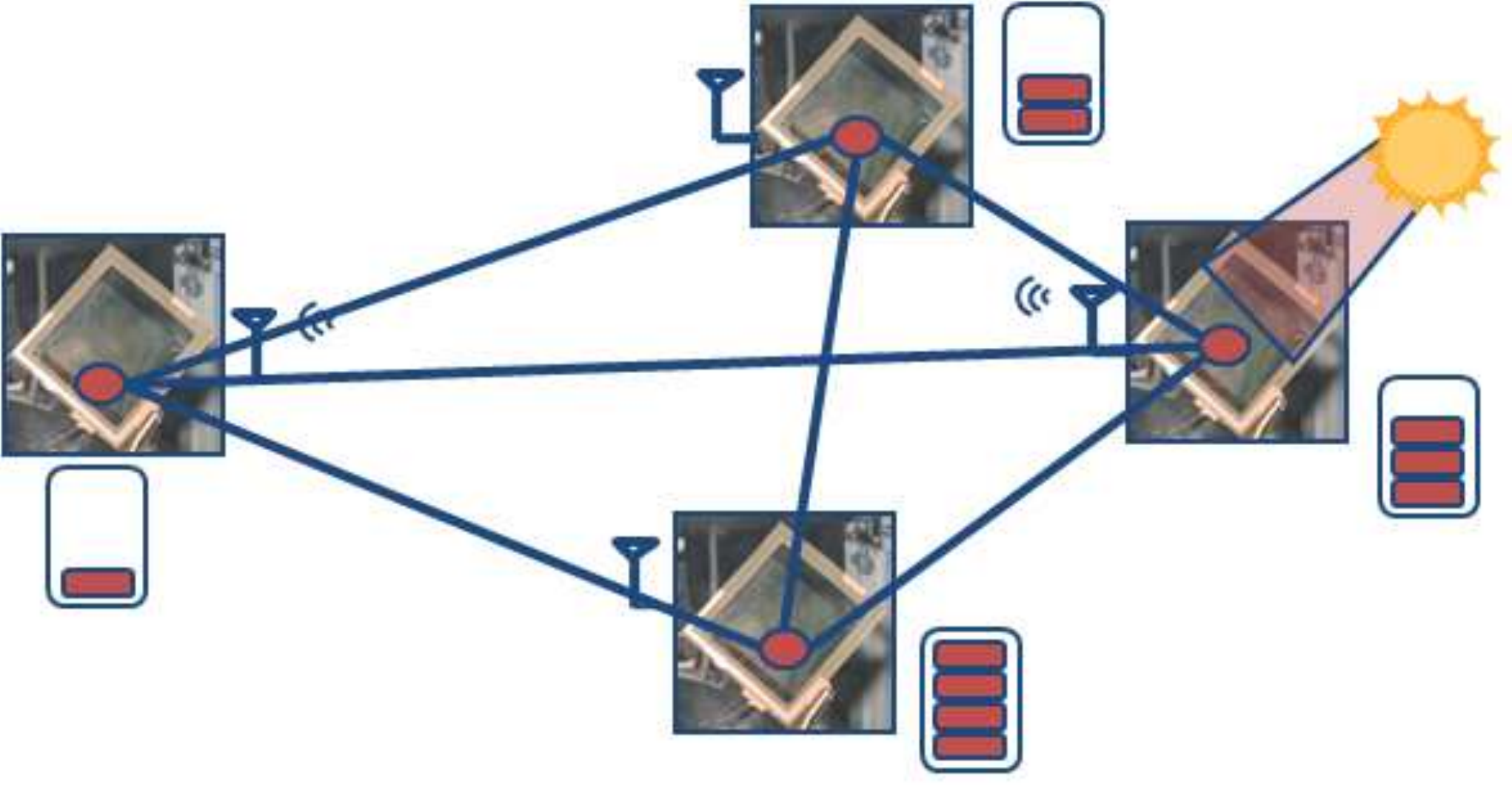}}
\vspace*{-0.3cm}
\caption{(a) The intended EnHANT form factor, (b) an EnHANT mockup integrated with a flexible solar cell and thin-film battery, and (c) an envisioned energy harvesting-adaptive EnHANT network. }
\end{figure}

The modern engineering landscape requires system engineering skills and interdisciplinary knowledge that are best acquired through participation in large-scale projects.
While the need to engage students in large-scale ``system perspective'' projects has been recognized~\cite{schocken2012taming,wolf2000embedded}, such projects are rarely attempted in academic settings. Project-based learning~\cite{lee2010project,hussmann2007crazy} is
actively used to help students build professional skills, such as teamwork and communication skills.
Despite the many project-based learning approaches that have been attempted and the many frameworks that have been proposed~\cite{Bernat2000structuring,wenderholm2004challenges,Coleman2012Collaboration,raicu2009enhancing}, to the best of our knowledge, project-based learning is typically only applied to small teams and small efforts.

In this paper we describe our experience in engaging a large and diverse group of students in project-based learning within a large-scale interdisciplinary research effort. Our experience with organizing multiple student projects to contribute to a large-scale effort is unique and this paper describes our approaches and some of the lessons we learned.

Our ongoing project-based learning activities are related to the \emph{``Internet of Things''}~\cite{Gorlatova_Enhants_wircom} -- digital networking of everyday objects.
Since 2009 a team of seven faculty members from the Department of Computer Science and the Department of Electrical Engineering at Co\-lum\-bia University have participated in the \emph{Energy Harvesting Active Networked Tags (EnHANTs)} project \cite{EnHANTsProject}.
The goal of the project is to develop a new type of a networked wireless device.
These small, flexible tags (the envisioned form factor for a future EnHANT is shown in Fig.~\ref{fig:mockup}, and a recent mock-up is shown in Fig.~\ref{fig:mckup2}) will be attached to commonplace objects to allow them to communicate and network.
EnHANTs will enable futuristic applications envisioned for the
Internet of Things, such as finding lost objects (i.e., lost keys, sunglasses, or toys will be accessible via wireless links) and
detecting object configurations. 

Recent advances in ultra-low-power wireless communications, energy harvesting (deriving energy from ambient sources such as light and motion), and energy harvesting-adaptive networking will enable the realization of EnHANTs in the near future~\cite{Gorlatova_Enhants_wircom}. Designing adaptive networks of EnHANTs, shown schematically in Fig.~\ref{fig:network}, requires reconsideration of protocols on all levels of the networking stack.
Additionally, designing the EnHANTs to achieve the desired form factor 
requires tight integration of the networking and communications protocols with the enabling hardware technologies. This necessitates close and continuous interactions of students and faculty members with expertise in the associated technology areas. Working on the EnHANTs project exposes the students to various disciplines in Electrical Engineering (circuit design, wireless communications), Computer Science (embedded systems, algorithm design, networking), and Applied Physics (battery and solar cell design).

Over 17~semesters, we have involved a diverse population of 80 high school, undergraduate, Masters, and Ph.D.~students in 180~semester-long research projects related to the design and development of the EnHANT prototypes and the prototype testbed.
The student projects are multidisciplinary. A project typically focuses on one disciplinary area, but requires interaction with at least two other areas. The projects necessitate collaboration, provide students with in-depth fundamental understanding of networking concepts, and require students to improve their communication skills. We use ``real-world'' system integration deadlines and frequent system demonstrations to motivate students and to encourage cross-disciplinary collaboration. Students demonstrated prototypes and the testbed at six conference demonstration sessions \cite{SeconDemoEnHANTs2010,MobiComDemo,SenSys2011Demo,MobiSys2011Demo,IDTechDemo2012,Margolies2013Demo} and at over three dozen additional live on-site and off-site demonstrations.
\makered{Students also contributed to publications describing the prototypes and the testbed~\cite{Gorlatova2013Prototyping,Gorlatova_EnHANTS_TOSN}.}
To evaluate our learning activities, we conducted a survey among the students.
Of the students who completed the survey, over 90\% indicated that the project was rewarding and enriching, and 70\% indicated that working on this project improved their ability to function on multidisciplinary teams \emph{more than any other activity in their academic career}.

This paper is organized as follows. Section~\ref{sect:Objectives} describes the
educational objectives and the related work.
Section~\ref{sect:Umbrella} describes the EnHANTs umbrella project and the student projects. 
Section~\ref{sect:Lessons} describes our approaches to organizing student projects and some lessons we learned over the 5-year course of the umbrella project.
Section~\ref{sect:Assessment} presents the evaluation results and Section~\ref{sect:outcomes} describes the outcomes of the project. Section~\ref{sect:Conclusion} concludes the paper.

\section{Context and Intended Outcomes}

\label{sect:Related}


Previous research has described structuring student research experiences as a course~\cite{polack2006learning,Greening2002Undergraduate}, a research-based program aimed at undergraduates~\cite{peckham2007increasing}, and a framework for accommodating undergraduate students in a research group~\cite{Bernat2000structuring,wenderholm2004challenges}. Researchers have also examined methods for providing students with interdisciplinary research opportunities ~\cite{raicu2009enhancing} and for increasing student communication and collaborative skills~\cite{Coleman2012Collaboration}. To the best of our knowledge, our experience with providing many students interdisciplinary project-based research opportunities as part of a single large-scale ongoing research effort is unique.

The EnHANTs research effort is in the embedded systems domain.
The applicability of project-based learning to embedded systems has been specifically noted. Additionally, the importance of learning embedded system design from a system perspective has previously been emphasized~\cite{wolf2000embedded}.
Project-based learning approaches for embedded system design have also been embraced in different courseworks~\cite{lee2010project,hussmann2007crazy}. The necessity of engaging students in large system development projects has been recognized as an important educational objective, and some tools for emulating the scale of the development have been proposed~\cite{schocken2012taming}.

\label{sect:Objectives}

\begin{figure}[t]
\centering
\subfigure[]{
\includegraphics[height=1.5in]{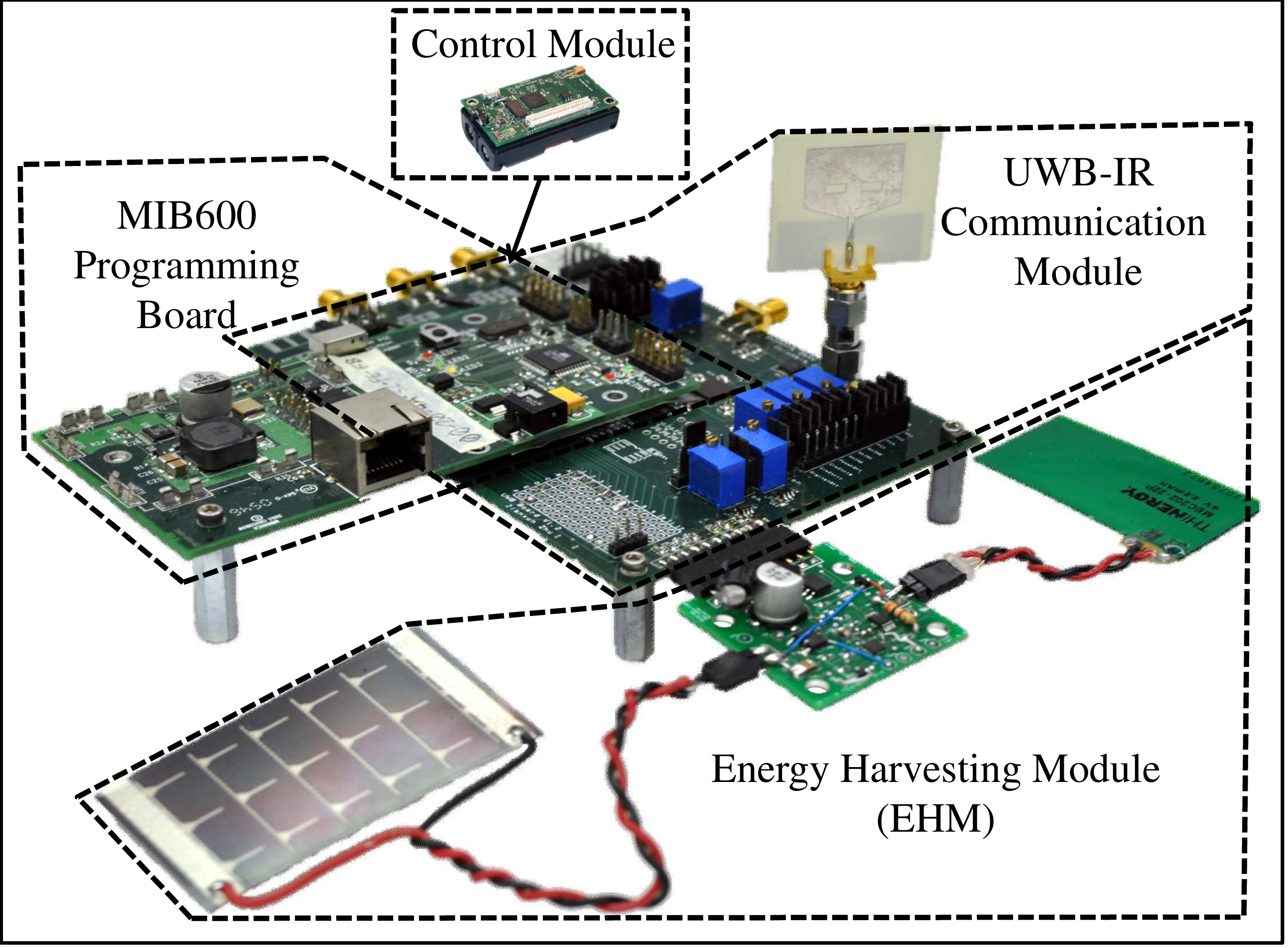}}
\subfigure[]{
\includegraphics[height = 1.5in]{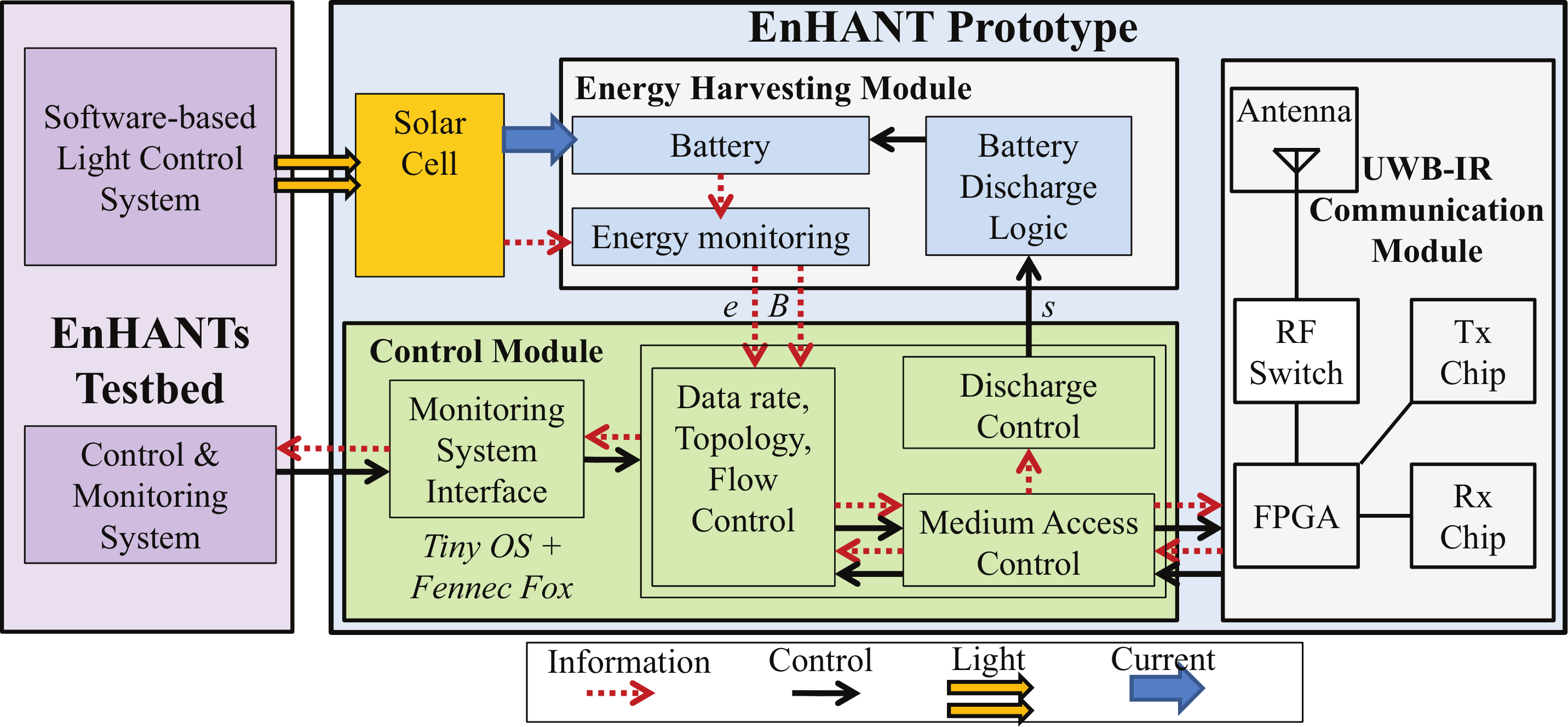}}
\vspace*{-0.35cm}
\caption{The EnHANT prototype: (a) photo, and (b) block diagram. \label{fig:protSchem}}
\end{figure}

This paper builds on the previous research by describing a method for engaging students in project-based learning within a large-scale multidisciplinary research effort, where the structure of the projects is critical to success \cite{saliklis2006putting}.
Project-based learning 
encourages system level thinking \cite{savage2007integrating}, helps develop engineering design skills \cite{dym2005engineering} and ``soft skills" \cite{hadim2002enhancing}, and is well-suited for team-based \cite{felder2000future} and long-term projects \cite{mills2003engineering}.
The research projects can help students to develop technical skills and encourage undergraduates to continue to graduate studies \cite{zydney2002impact}.

\begin{figure*}[t]
\centering
\centering
\includegraphics[width=1.9\columnwidth]{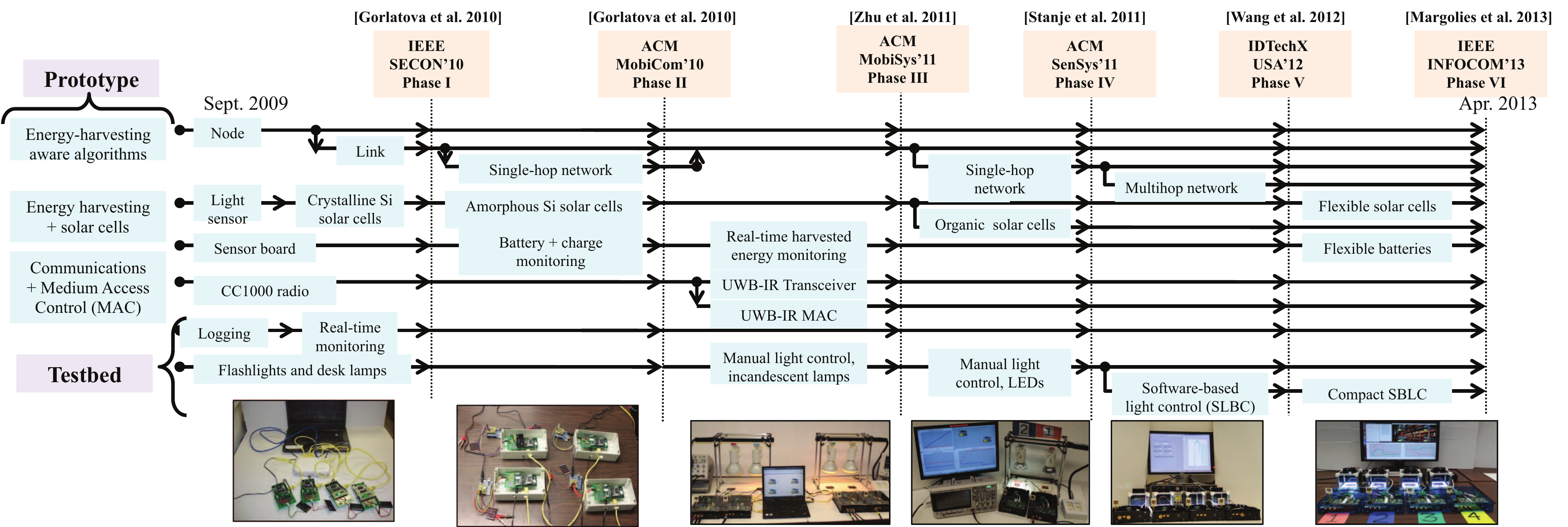}
\caption{A timeline of the EnHANTs prototypes and testbed development,
  Phases I-VI. The milestones achieved during these phases were presented at conference
  demonstration sessions \cite{SeconDemoEnHANTs2010} (7 student coauthors),
  \cite{MobiComDemo} (7 student coauthors), \cite{MobiSys2011Demo} (9 student coauthors),
  \cite{SenSys2011Demo} (10 student coauthors),
  \cite{IDTechDemo2012} (10 student coauthors), and \cite{Margolies2013Demo} (8 student coauthors).
  \label{fig:PrototypeDevelopmentStages}\label{fig:PhasesPhoto}}
\end{figure*}

The high-level educational goal of the project is to improve the student's overall education by exposing them to a large-scale multidisciplinary research project. The project's main educational objectives focus on improving the students' abilities to function on multidisciplinary teams and communicate effectively. Specifically, some notable objectives were to enable students to effectively understand a \emph{systems} view of their specific project in the context of the overarching umbrella project. It was also our goal to enable students to effectively communicate with team members of different backgrounds (i.e., hardware and software) as part of a unique multidisciplinary design and development process.
Other learning outcomes, some of which were inspired by ABET EC 2000~\cite{ABET2000} included improving the students ability to design and conduct experiments and analyze and interpret data. In Section~\ref{sect:Assessment} we discuss evaluation results related to some of these objectives. 



\section{Project-based Learning} \label{sect:Umbrella}

The Energy Harvesting Active Networked Tags (EnHANTs) umbrella proj\-ect~\cite{EnHANTsProject} has engaged, over 17~semesters, 80~students in 180 interdisciplinary hands-on semester-long projects. 
In this section, we first briefly describe the umbrella project.
Then, we describe the student involvement.

\subsection{Umbrella Project}


The EnHANTs proj\-ect~\cite{EnHANTsProject} is a large multi-year interdisciplinary research project related to the \emph{Internet Of Things}. The main focus of the project is the development of the EnHANTs prototypes and the EnHANT prototype testbed. 
The current prototype is shown in Fig.~\ref{fig:protSchem}.
The prototypes harvest indoor light energy using thin-film solar cells, communicate with each other using \mbox{custom-developed} 
transceivers, and implement custom energy-harve\-sting-adaptive 
algorithms on several layers of the Open Systems Interconnection 
stack. 
The testbed 
monitors the prototypes' communications and networking parameters and controls the amount of light incident on the prototypes' solar cells.

The prototypes and the testbed were developed in a set of phases over 5~years. The photos of the testbed in the different phases are shown in Fig.~\ref{fig:PhasesPhoto}, which additionally demonstrates the iterative development of the different aspects of the prototype and testbed functionality. We summarize the functionality development steps below. 
\emph{All hardware, software, and algorithm modifications to the prototype and testbed functionality
were designed, developed, integrated, and tested as part of student projects}. 
\begin{myitemize}
\item \textbf{Energy harvesting} allows EnHANTs to self-power by obtaining energy from ambient sources (light, motion).
    As part of the umbrella project, we have fabricated \emph{flexible solar cells} that
    efficiently harvest indoor light, and integrated them with the prototypes.
    Initially, we designed the prototypes to sense, but not harvest, available environmental energy (Phase~I).
    Next, we integrated rigid commercial solar cells and implemented real-time energy harvesting state monitoring (Phases~II and III). Finally, we integrated the custom-designed organic solar cells (Phase~V) and commercial flexible solar cells (Phase~VI). 
\item \textbf{Ultra-Wideband \mbox{Impulse-Radio} (\mbox{UWB-IR}) wireless communications}
    spend significantly less energy than other low-power wireless technologies~\cite{Crepaldi2011}.
    Early-phase prototypes communicated with each other via standard (\mbox{non-UWB}) commercial sensor network mote transceivers~\cite{MicaMote}. Prior to integration of the custom \mbox{UWB-IR} communication modules
    in Phase~III, we substantially modified the mote operating system (which did not support custom transceivers).
    The integration additionally required the implementation of a custom medium access control module, since the \mbox{UWB-IR} transceiver characteristics differ greatly from the properties of the conventional transceivers. 
\item \textbf{Energy harvesting-adaptive algorithms} were first designed and developed 
    for simple single node scenarios, and were later implemented for network scenarios.
    Following the integration of the \mbox{UWB-IR} transceivers in Phase~III, we
     re-implemented the algorithms to take the \mbox{UWB-IR}
     characteristics into account. In Phase~VI, we introduced an adaptive multihop network.
\item \textbf{Testbed functionality} first consisted of a data logger with a simple visualization interface, which we  
    replaced with a custom-designed real-time monitoring and control system. 
    We additionally 
    developed several prototype light energy control systems,
    from relatively simple manual setups 
    (Phases~III and IV) to a software-based system that exposes the prototypes to real-world trace-based light energy conditions (Phase~V).
\end{myitemize}

At the end of each phase, a fully functional prototype testbed was demonstrated by students, as indicated in Fig.~\ref{fig:PhasesPhoto}:
\begin{myitemize}
\item Phase~I demo at IEEE SECON'10~\cite{SeconDemoEnHANTs2010}: 7 student coauthors from 2 research groups.
\item Phase~II demo at ACM MobiCom'10~\cite{MobiComDemo}: 7 student coauthors from 4 research groups.
\item Phase~III demo at ACM MobiSys'11~\cite{MobiSys2011Demo}: 9 student coauthors from 4 research groups.
\item Phase~IV demo at ACM SenSys'11~\cite{SenSys2011Demo}: 10 student coauthors from 4 research groups.
\item Phase~V demo at IDTechX'12 Energy Harvesting and Storage USA Conference~\cite{IDTechDemo2012}: 10 student coauthors from 3 research groups.
\item Phase~VI demo at IEEE INFOCOM'13~\cite{Margolies2013Demo}: 8 student coauthors from 3 research groups.
\end{myitemize} 
The Phase~IV demo~\cite{SenSys2011Demo} 
received the conference's \emph{Best Student Demonstration Award}.

The conference demonstrations were conducted in different cities, which necessitated solving not only technological but also logistical challenges. For example, Phase II, IV, and VI demonstrations required shipping. The Phase~VI demonstration~\cite{Margolies2013Demo} was conducted abroad, which exposed students to the differences in electrical systems of different countries and to some of the challenges of international shipping logistics.

\makered{Additionally, we recently described the prototypes and the testbed in two publications~\cite{Gorlatova2013Prototyping,Gorlatova_EnHANTS_TOSN} (7 student coauthors from 4 research groups;
16 contributing students from 5 research groups additionally acknowledged).}

\subsection{Student Projects} \label{sect:StudentProjects}

\begin{figure*}[t]
\centering
\adjustbox{width=1.4\columnwidth}{\trimbox{1cm 0cm 1cm 0cm}{\includegraphics[width=1\columnwidth]{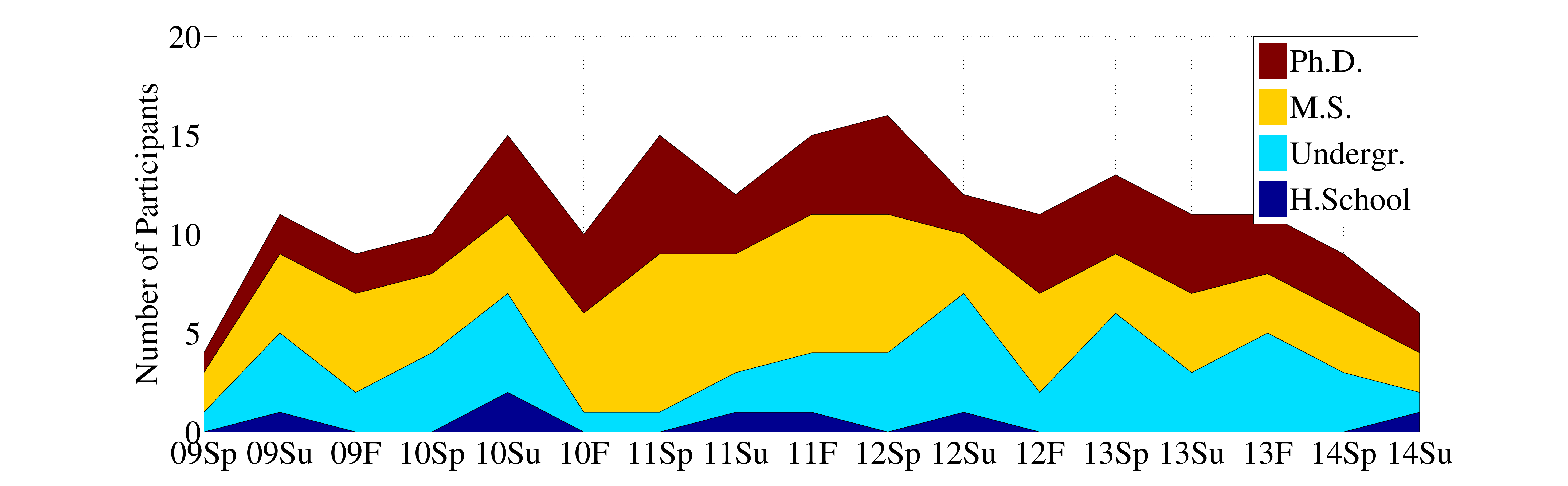}}}
\vspace*{-0.2cm}
\caption{\label{fig:studentsOverTime} Student projects by semester.}
\end{figure*}

\begin{figure}[t]
\centering
\subfigure[Academic level.]
{\includegraphics[width=0.45\columnwidth]{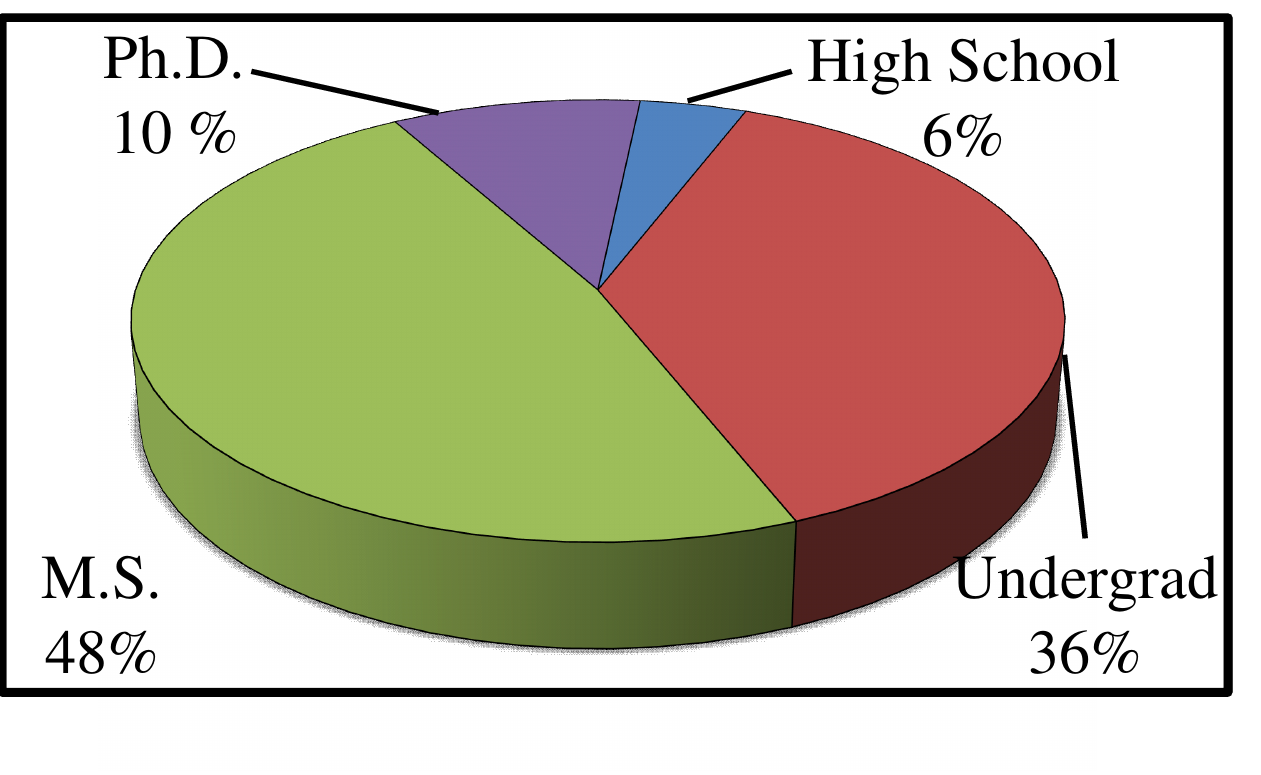}}
\subfigure[Academic affiliation.]
{\includegraphics[width=0.45\columnwidth]{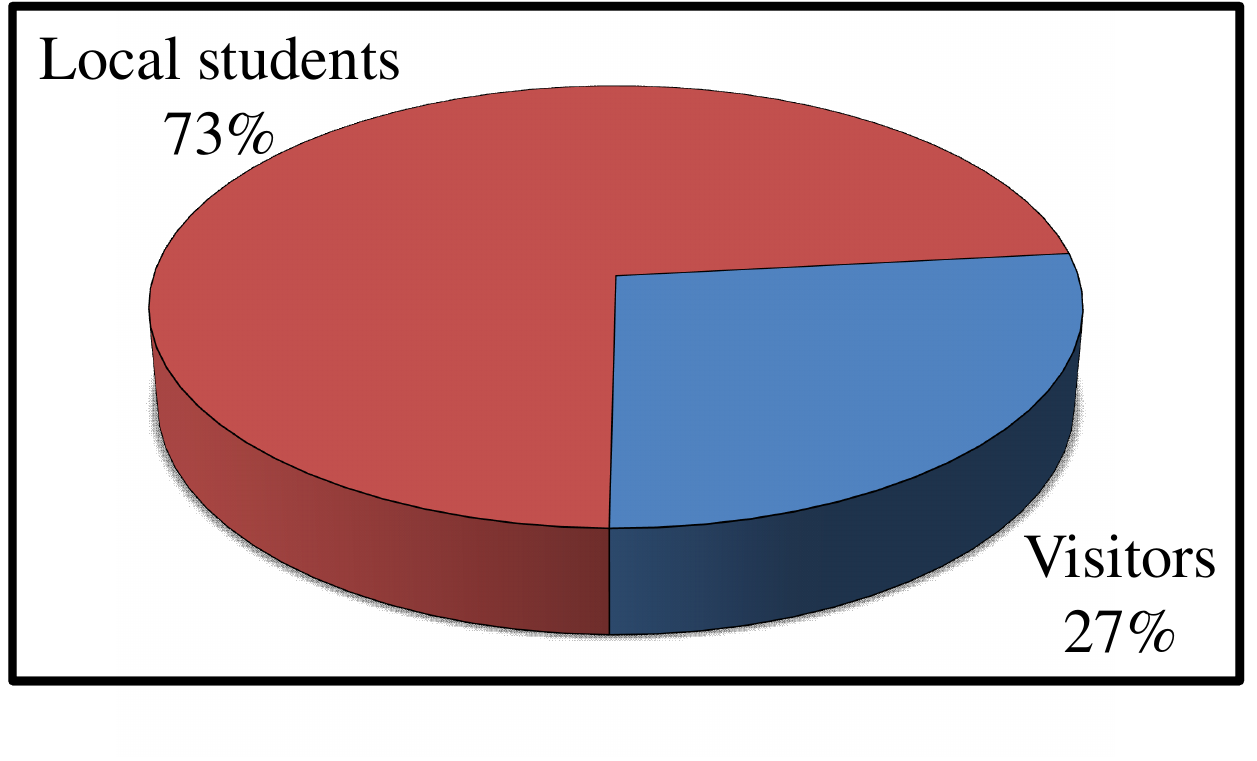}}
\vspace*{-0.3cm}
\caption{\label{fig:studentBreakdown}Students involved in the EnHANTs project.}
\centering
\end{figure}

\begin{figure}[t]
\subfigure[Gender.]
{\includegraphics[width=0.45\columnwidth]{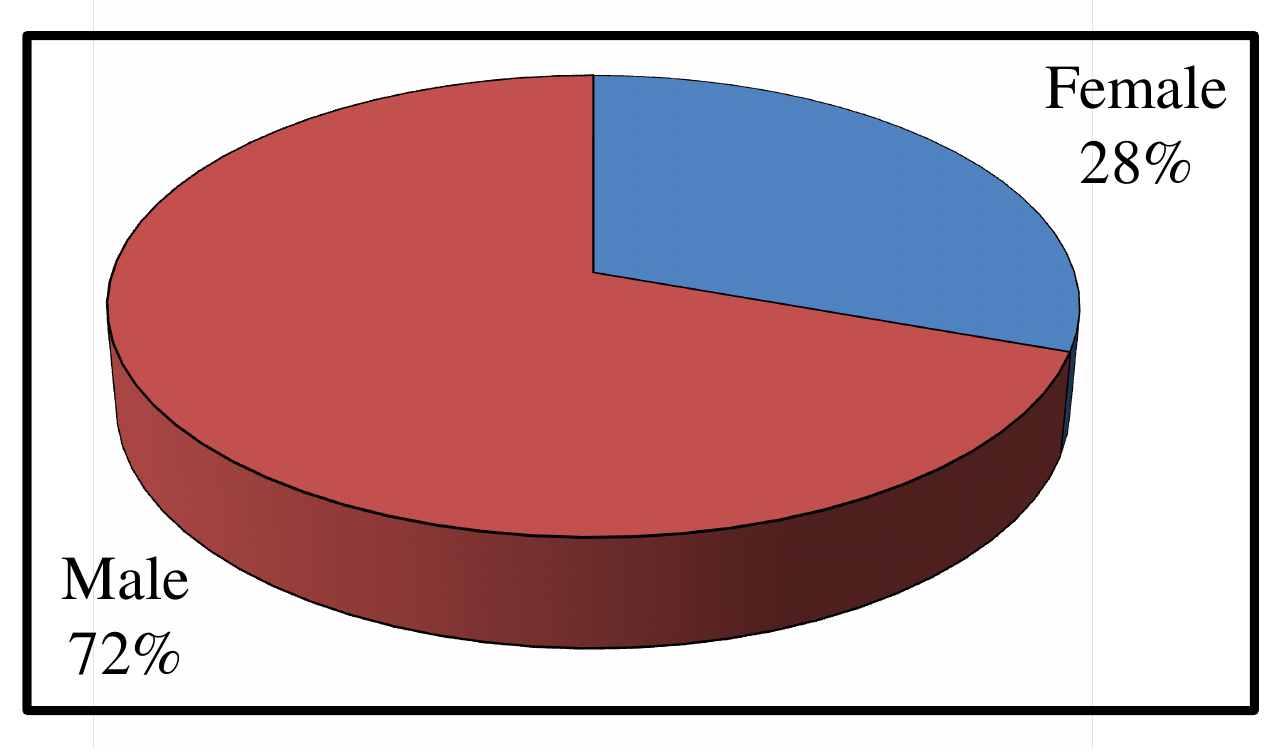}}
\subfigure[Main focus area.]{
\includegraphics[width=0.45\columnwidth]{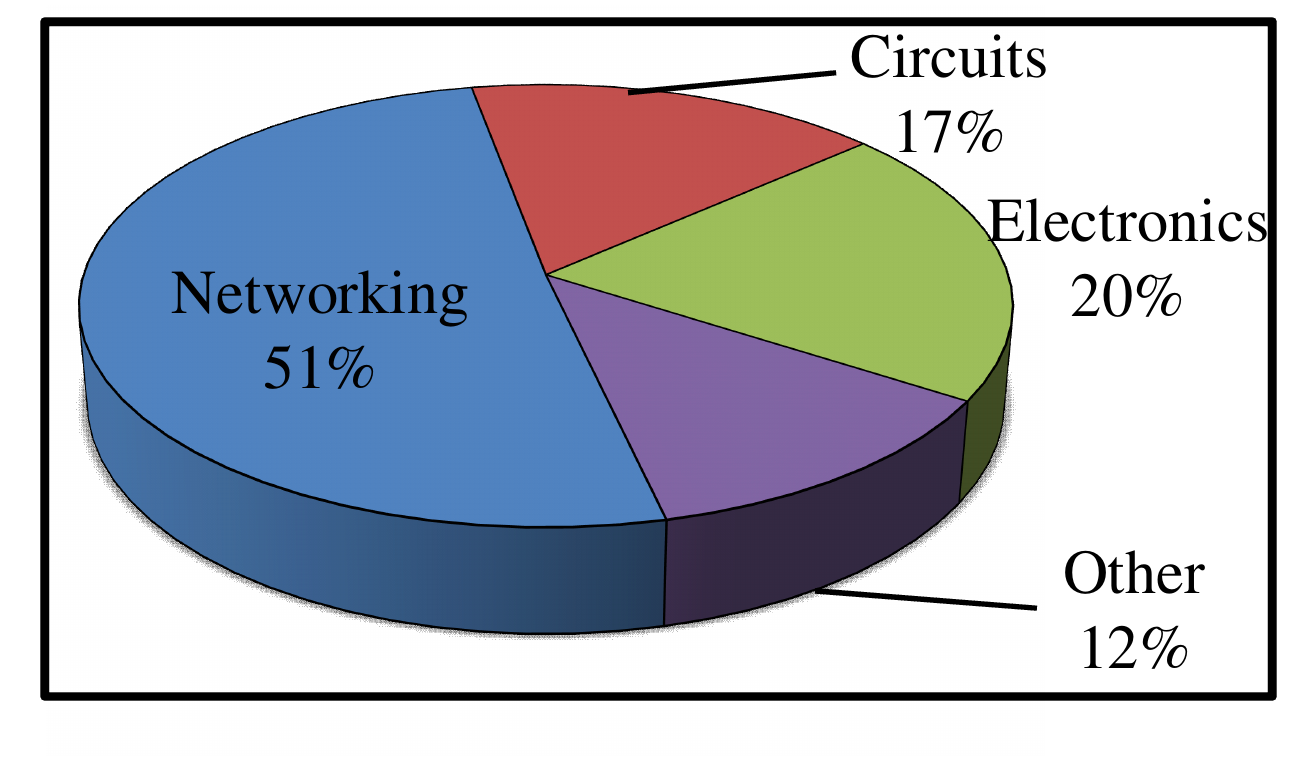}}
\vspace*{-0.3cm}
\caption{\label{fig:studentSemesterBreakdown}Students semesters.}
\end{figure}

As of September 2014, over 17~semesters the EnHANTs umbrella project 
has engaged 80~students who have completed 180~semester-long projects.
The number of student projects completed each semester is shown in Fig.~\ref{fig:studentsOverTime}. The student demographics are presented
in Fig.~\ref{fig:studentBreakdown} and Fig.~\ref{fig:studentSemesterBreakdown}.
Out of 80 students, 6\% were high school students we engaged via a Harlem Children Society's program that pairs university researchers with students from under-served communities~\cite{HarlemChildrenSociety}. 36\% were undergraduates, 48\% were M.S. students (nearly all of which were in non-thesis terminal M.S. programs), and 10\% were Ph.D.~students. 73\% of students were enrolled in academic programs in Columbia University, while the other 27\% were visiting students, such as Research Experience for Undergraduate (REU) students, students from local colleges without advanced research facilities, or visiting international students. \makered{Students that were enrolled in academic programs at Columbia University typically took this project as an elective independent research project course (separate from the senior design capstone project)}.
Out of the 180~student projects, 51\% 
were full-time projects (summer research internships, REU projects, M.S.\ thesis research semesters).
The other 49\% were semester-long research project courses to which students typically dedicated 8-15 hours per week. 72\% of the projects were completed by male students and 28\% by female students. The main focus areas for the student projects were networking (51\%), circuits and systems (17\%), and electronics and applied physics (20\%). Other projects focused on operating systems and VLSI design (12\%).

The student projects within the EnHANTs project are collaborative and multidisciplinary.
A project typically focuses on one disciplinary area (e.g., algorithm design, operating systems development, solar cell design) 
but requires interaction with at least two other areas. These projects challenge students by requiring them to gain understanding of concepts outside of their comfort zone. Additionally, the projects require students to independently and proactively seek out relevant expertise throughout the research groups involved in the EnHANTs project. This improves students' communication and teamwork skills.
Four representative student projects are described below. More than 40 additional project descriptions are provided at~\cite{EnHANTsProject}. 

\begin{myitemize}

\item \emph{Real-time monitoring and control system} (Phase~I project, completed by an undergraduate Computer Science student):
The student developed a \mbox{Java-based} system to monitor and control the EnHANTs prototypes. The project involved designing the necessary data structures to enable communication between the prototypes and the computer running the monitoring system. The student designed the system to support both a text-based interface and a ``visual demo'' interface that shows the activity of the prototypes in an easy-to-understand way. This project, implemented using TinyOS and Java, required knowledge of sensor networking, wireline communication, and software design. The student extensively interacted with students who were modifying the prototype operating system and developing energy-harvesting-aware algorithms.

\item \emph{Energy harvesting module} (Phase III project, completed by an M.S. Electrical Engineering student): The student improved the design of the Energy Harvesting Module (EHM) that tracked the energy harvested from the solar cell and the state of the battery in real time. The project required the student to improve on a simple prototype that used an off the shelf Coulomb counter. The student worked with the students developing energy-harvesting-adaptive algorithms to define a new specification for the range, resolution, and response time of the EHM. The student designed and built a new EHM that included a high side current sense amplifier and new digital interface to meet these requirements. This project required knowledge of circuit design and embedded programming and an understanding of how the underlying hardware would affect higher level system performance.

\item \emph{Prototype UWB-IR communication module} (Phase~III proj\-ect, completed by an  M.S.\ Electrical Engineering student):
    The student developed and tested the \mbox{UWB-IR} communication module. The student integrated a custom-designed \mbox{UWB-IR} transmitter and receiver chipset onto a single printed circuit board and programmed a complex programmable logic device     to perform data serialization and deserialization, preamble detection, and byte synchronization. The student also developed a UWB-IR radio driver using TinyOS. Primarily focused on circuit design, this project required the student to develop expertise in networking, operating systems, and software design.

\item \emph{Energy harvesting-adaptive 
    network} (Phase~V proj\-ect, completed by an undergraduate Computer Engineering student): The student implemented network layer protocols that handled the EnHANT packet routing.
    The student first tested the network functionality using the commercial transceivers, and then
    extensively evaluated its performance with the custom \mbox{UWB-IR} transceivers.
    The student also implemented energy harvesting-adaptive network layer algorithms, which changed packet routing paths based on the environmental energy availability. The student extensively tested these algorithms with the custom energy harvesting modules.
    While primarily focused on networking, the project required the student to gain an in-depth knowledge of energy harvesting and \mbox{UWR-IR} communications.
\end{myitemize}

\section{Project Organization and Lessons Learned}
\label{sect:Approach}\label{sect:Lessons}
Organizing multiple student projects to contribute to a large-scale effort is challenging.
In this section we describe how the 
student projects were structured to facilitate student learning. 
We also present lessons that were learned throughout the \mbox{5-year} course of the project.
The lessons learned focus on \emph{real-world experiences} including close and continuous cross-group interactions, well-defined interfaces between student projects, structured short-term goals, real-world work environment, development of student mentors, and project self-promotion.

\subsection{Multidisciplinary Projects}
As described in Section~\ref{sect:Umbrella}, the student projects are collaborative and multidisciplinary, often requiring interaction with at least two technical areas. The students work in different labs, focus on different disciplines, and have different technical skills, priorities, work styles, and expectations.

Early on we discovered that the gaps between the knowledge of the students with different expertise areas are much wider than anticipated. For example, Electrical Engineering students are oftentimes unfamiliar with good software development practices, while many Computer Science students may not understand how to handle experimental electronics. Additionally, many students that are not majoring in Electrical Engineering do not understand the concepts of frequency-domain signal processing that are essential to the understanding of wireless networking.

These knowledge gaps often lead to both technological and interpersonal issues. Cross-group problem solving requires students to trust each other's expertise, but these gaps in knowledge can make the trust difficult to establish.
When working with students, we encourage close and continuous cross-group interactions to address these issues.
We highlight that such gaps are normal and should be treated as learning opportunities. 

Additionally, some of the most challenging problems arise when a student is interfacing his or her project with another project. The difficulty of solving these interface problems may lead to interpersonal tensions.
In our experience, designing well-defined interfaces between different technologies has been a challenging task that often required faculty members' involvement. Ultimately these issues can only be addressed by
establishing, maintaining, and nurturing the connections between the research groups and between the different students.

\subsection{Real-world System Integration Deadlines}

EnHANTs prototype and testbed design, development, and integration have proceeded in a series of phases (see Fig.~\ref{fig:PrototypeDevelopmentStages}). At the end of each phase, the fully integrated prototype and testbed were presented at a major conference.

Using conference timelines as real-world deadlines for the integration of different student projects has many benefits. Providing structured short-term deadlines for student projects, rather than abstract long-term goals, energizes and motivates the students. Students are additionally motivated by seeing their work integrated with the work of others, used in conference presentations, and subsequently extended.
We learned that short-term system integration deadlines also encourage frequent cross-disciplinary collaboration, as the students worked together to quickly solve problems that were arising. In addition, constant updates to the software and the hardware throughout the integration deadlines reduce the impact of unsuccessful projects.

\begin{figure*}[t]
\centering
\includegraphics[height=2.1in]{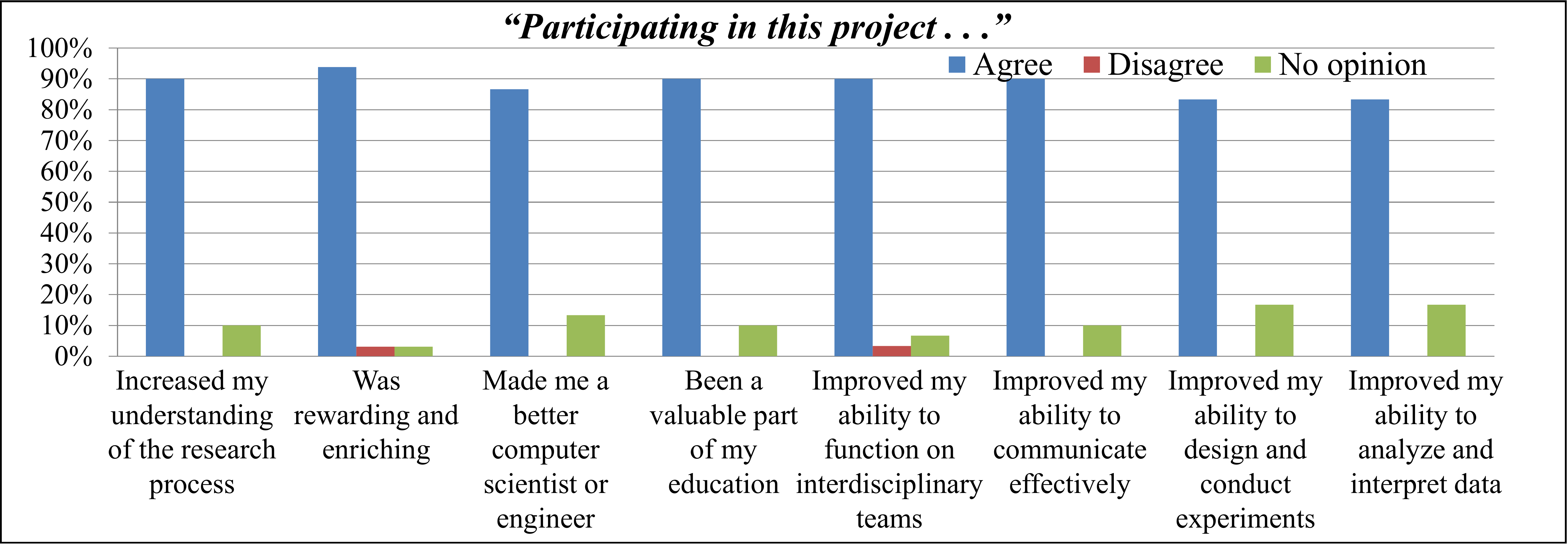}
\caption{ \label{fig:SurveyHighlights} Project survey results: Effects of project participation on educational objectives.}
\end{figure*}

\subsection{Frequent Cross-group Meetings}

Students present their work to the faculty and students from the different research groups at regular (weekly or bi-weekly) meetings. Such frequent cross-group meetings resemble interdepartmental meetings that are used to manage large-scale projects in industry environments. The frequent meetings facilitate cross-group problem-solving and encourage new team members to quickly get up to speed on the state of the projects. Additionally, the presence of faculty and students from different research groups at these meetings challenges students to present their work so that it could be understood by people with different backgrounds, enabling multidisciplinary discussions. The students also report that observing how faculty members solve problems during these meetings improves their own problem solving skills.

\subsection{Frequent System Demonstrations} 
\begin{figure}[t]
\begin{minipage}{0.4\columnwidth} 
\centering
\includegraphics[height = 1in]{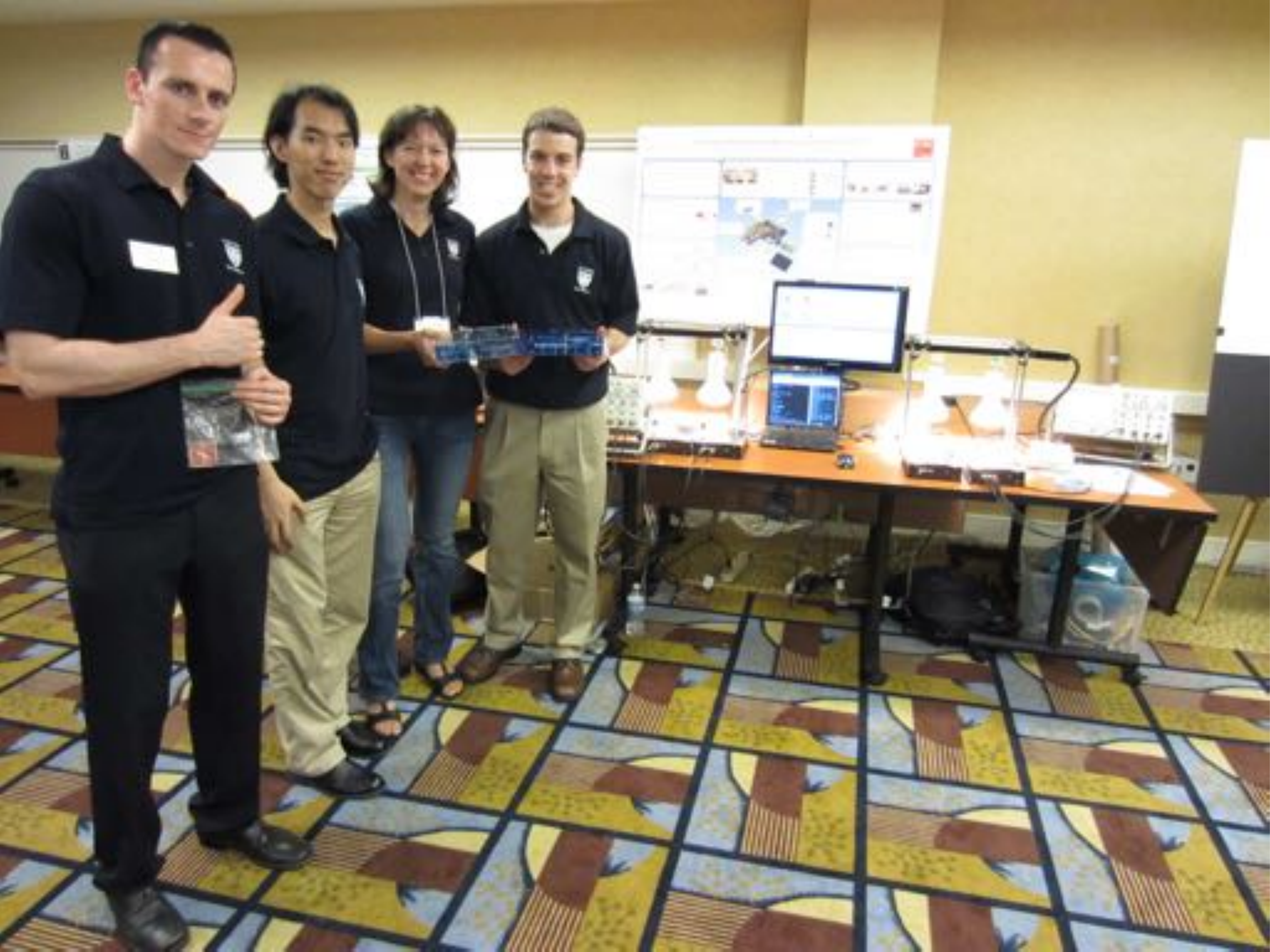}
\caption{A Phase III conference demonstration~\cite{MobiSys2011Demo} presented by M.S.\ and Ph.D.\ students.\label{fig:MobisysDemo}}
\end{minipage}
\hspace*{0.08\columnwidth}
\begin{minipage}{0.4\columnwidth}
\centering
\includegraphics[height=01in]{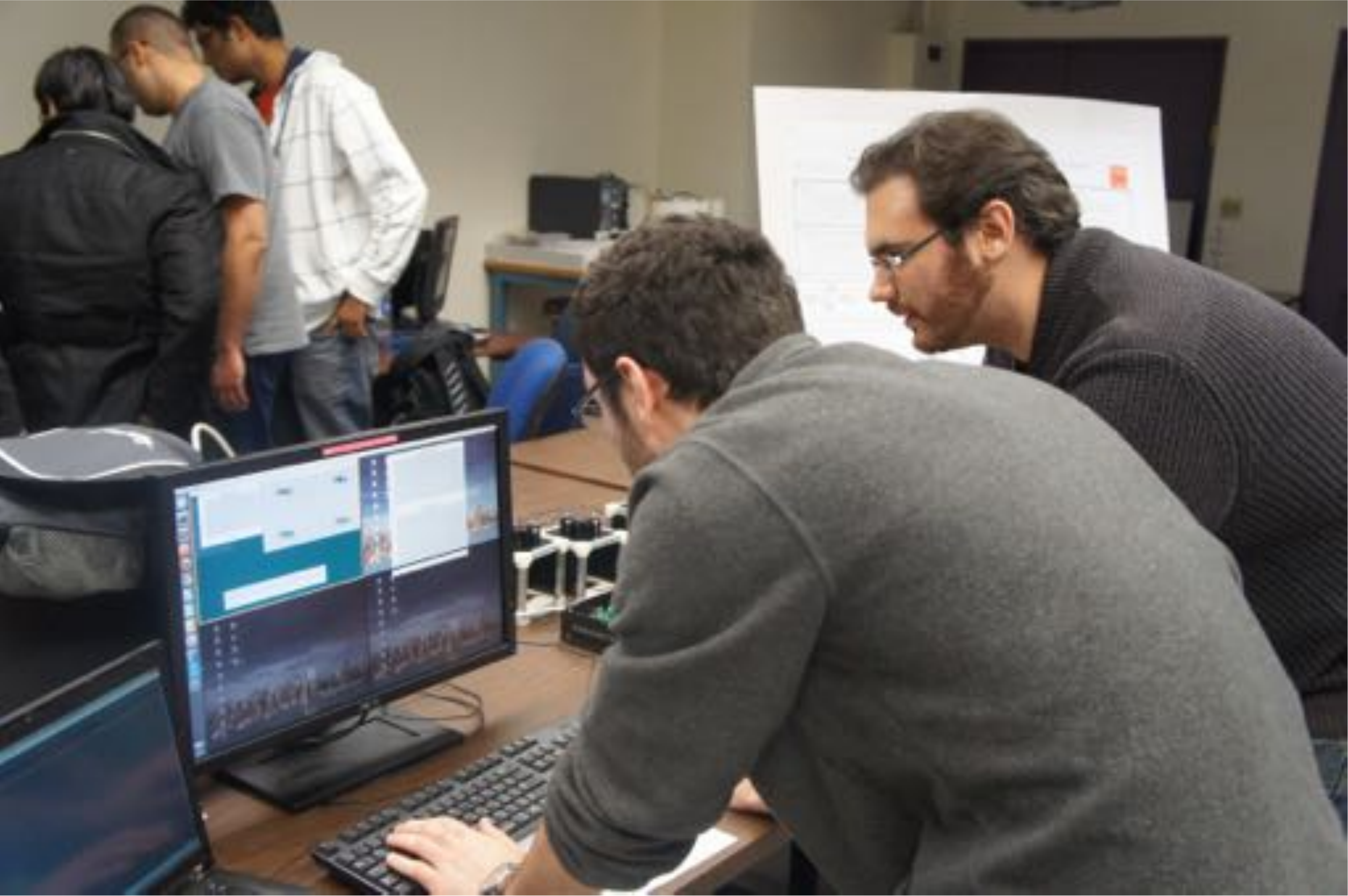}
\centering
\caption{\label{fig:Mentoring} A Ph.D.\ student mentoring an undergraduate student.\\ }
\end{minipage}
\end{figure}

Functional ``live'' EnHANT prototypes and testbed are frequently demonstrated in different on-site and off-site presentations.\footnote{Inspired by agile software development practices, we ensure that a version of the prototypes and testbed is always ready to be demonstrated. We do not integrate new software or hardware without extensive testing and design for backward compatibility.
This further reduces the impact of unsuccessful student projects.}
Frequent demonstrations, particularly those conducted off-site (for example, Fig.~\ref{fig:MobisysDemo} shows the Phase~III conference demonstration setup assembled at the conference location by the student presenters), encourage students to design and develop robust software, hardware, and algorithms and to extensively verify and test their work. This improves students' technical skills, and provides them with an understanding of the quality standards required from technology in ``real-world'' applications. Additionally, the testbed demonstrations give students opportunities to present their work to vastly different audiences.

\makered{The frequent demonstrations of the EnHANTs system necessitate rigorous risk management. For example, we prepare and fully test 2--3 ``hot'' backup components for each of the demonstration components. The backups are tested both as individual components and while integrated with the rest of the system. 
Off-site demonstrations entail rigourous contingency planning. Included in the many logistics to be considered are international shipping of our testbed, prototypes, and monitoring system and voltage and plug adapters to accommodate international power supplies. }

\subsection{Ph.D.~Student Mentorship}
The faculty members involved in the EnHANTs project are heavily engaged in the student projects.
However, faculty members delegate many of the day-to-day student supervision tasks to the Ph.D.~students.
The Ph.D.\ students provide technical support and guidance to the students (for example, Fig.~\ref{fig:Mentoring} shows a Ph.D.\ student troubleshooting a prototype component together with an undergraduate student). Additionally, the Ph.D.\ students are responsible for testing and
verifying junior student projects before integration with EnHANTs prototypes and testbed, and for ensuring continuity between the different student projects. 

\makered{Handling many of the day-to-day activities, the Ph.D.\ students quickly learn student mentorship skills from the faculty to become advisors to the high school, undergraduate, and M.S. students.
This awards the Ph.D.\ students with opportunities to practice teaching, assigning deadlines, and motivating their advisees. 
Additionally, the Ph.D.\ students serve as initial reviewers for all presentation materials prepared by the junior students, which allows the Ph.D.\ students to practice giving feedback (both technical and related to different aspects of written and verbal communications). This experience provides the Ph.D.\ students with insights into the managerial aspects of academic positions, as well as with high level skills that are valuable in industry environments.} 

\subsection{Outreach and Promotion} \comment{new title here?}

\begin{figure}
\centering
{\includegraphics[width=0.45\columnwidth]{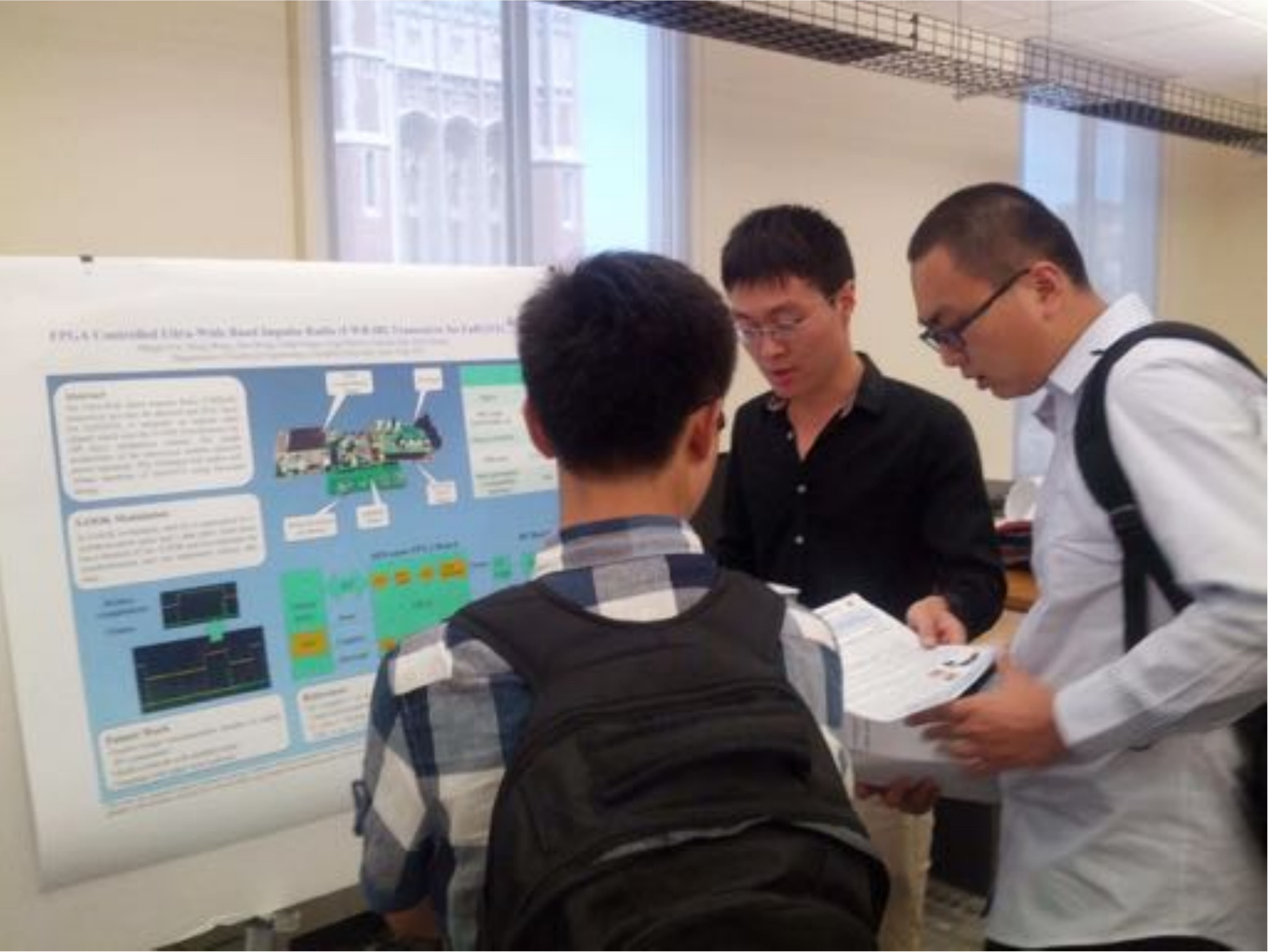}}
\hspace*{0.03\columnwidth}
{\includegraphics[width=0.45\columnwidth]{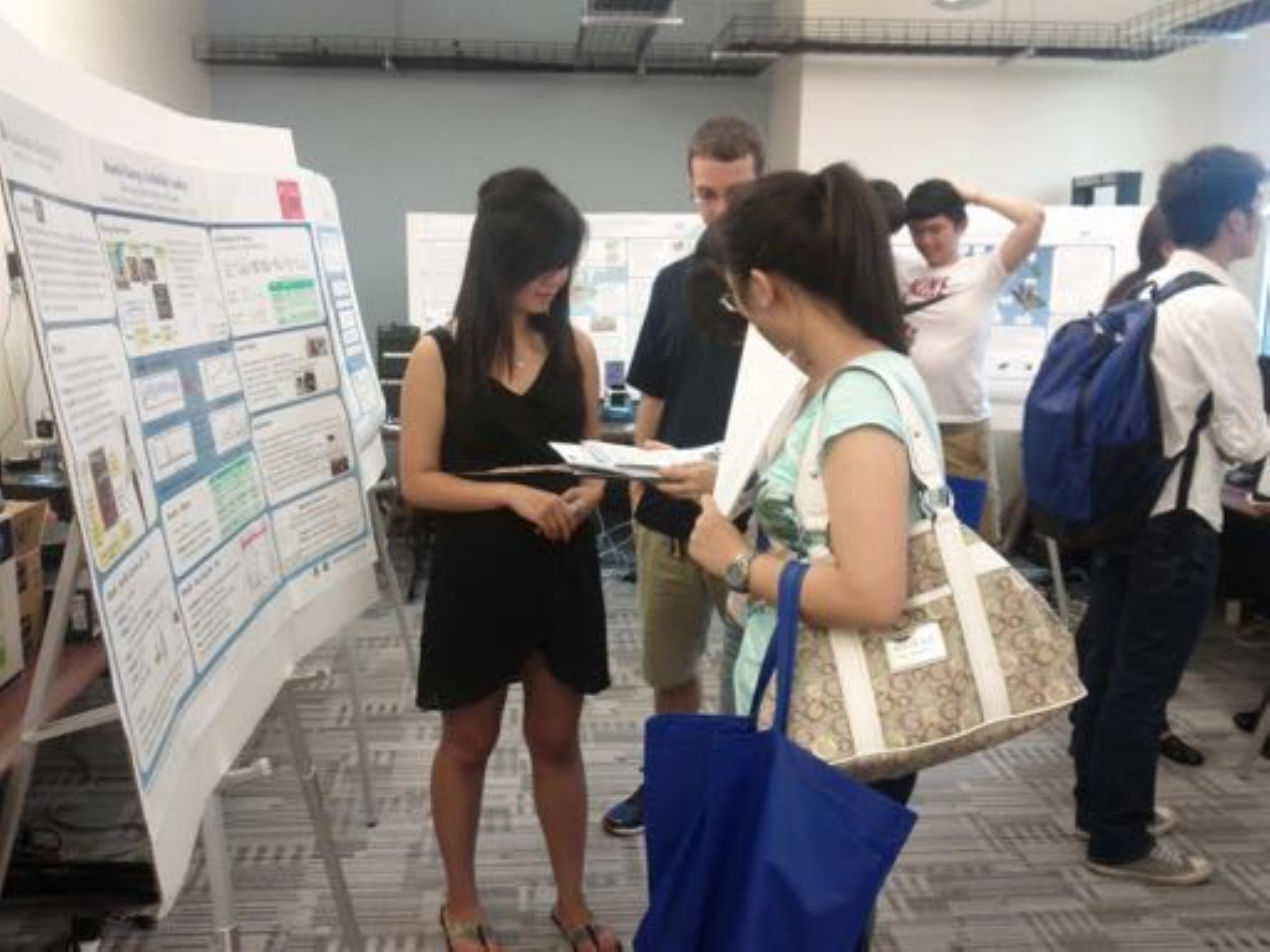}}
\caption{\label{fig:MSPosterSession} Undergraduate and M.S.\ students presenting their work to incoming M.S.\ students in Summer 2013. }
\end{figure}

\makered{Students also serve as ambassadors for the umbrella project,
performing tasks that are oftentimes typically done by 
faculty. For example, the students often directly present their research to
prospective graduate and undergraduate students (see Fig.~\ref{fig:MSPosterSession}) and to university visitors. Due to the large scale of the project, the students
regularly get opportunities to demonstrate their work to high-profile visitors
they would not otherwise encounter. This encourages the students to take pride in showcasing their work.
Among the visitors with whom the projects' undergraduate and M.S.\ students interacted are
many top academic researchers from the networking, communications, and electronics communities.
Additionally, many visitors came from different companies including IBM, Juniper, Samsung, and Phillips. 
The students interacted with technical and non-technical company representatives. 
The non-technical representatives included visitors with roles in marketing, operations, and sales.}

\begin{figure}
\centering
\includegraphics[width=0.8\columnwidth]{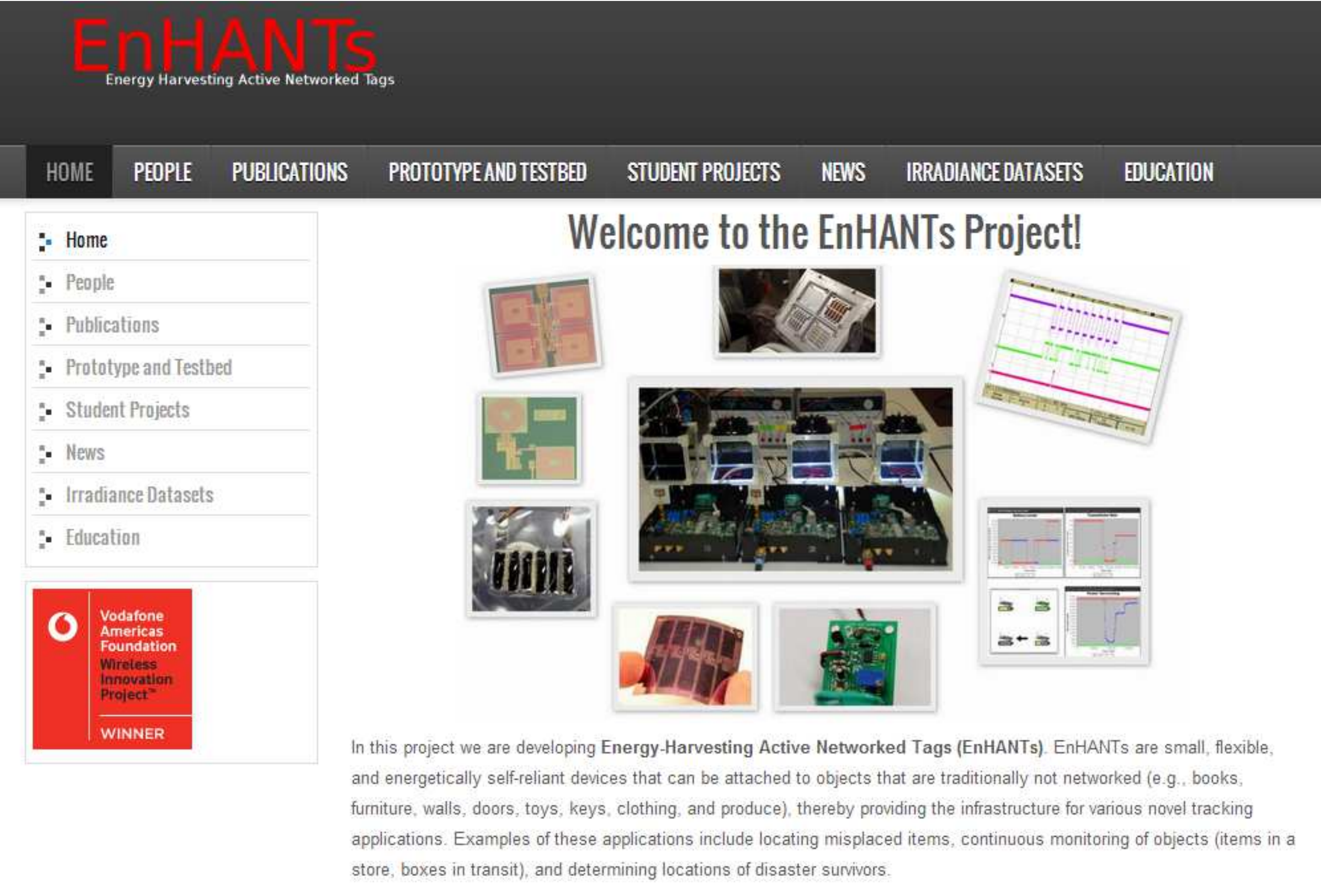}
\caption{\label{fig:website}A screenshot of the umbrella project website welcome page~\cite{EnHANTsProject}. }
\end{figure}

\makered{Additionally, the students publicize their contributions
to the project on the EnHANTs website (see \cite{EnHANTsProject} and Fig.~\ref{fig:website}), which is maintained by the members of the project.
At the end of each project, the students publish the results of their work on the website. This process encourages students to step back from the intensity of their research,
see their work in a positive light, and present it in an exciting manner.
We learned that providing students with venues for displaying their work 
motivated students.  
The direct involvement of students in the project promotion activities was also useful in recruiting new students. }

\subsection{Knowledge Transfer}

The large-scale, long-term nature of the umbrella project necessitates knowledge transfer between the students. The project uses an internal wiki that is kept up to date with shared technical information and documentation. We learned that, while knowledge transfer needs to be carefully monitored and emphasized, most students embrace it when they can see first-hand that the documentation they create is used by their peers and mentors. Similarly, most students embrace the opportunity to introduce peers to their work and to teach them.

\section{Experiences and Feedback}
\label{sect:Assessment}
%

After the completion of Phase~V of the umbrella project in October 2012 (see Section~\ref{sect:Umbrella}), we conducted a survey to evaluate student learning and the effectiveness of our approaches. In this section, we summarize the results of the survey and other student feedback and briefly discuss our changes to the organizational structure of the umbrella project based on the feedback. We also briefly summarize student outcomes.

We conducted a survey among all 45~high school, undergraduate, and M.S.~students that had participated in the project up to October 2012. The survey contained multiple-choice questions 
and optional open-ended questions. Answers to the multiple-choice questions were measured on a Likert scale:
1 -- strongly disagree, 2 -- disagree, 3 -- neither agree nor disagree, 4 -- agree, 5 -- strongly agree.

\begin{figure}
\centering
\includegraphics[width=\columnwidth]{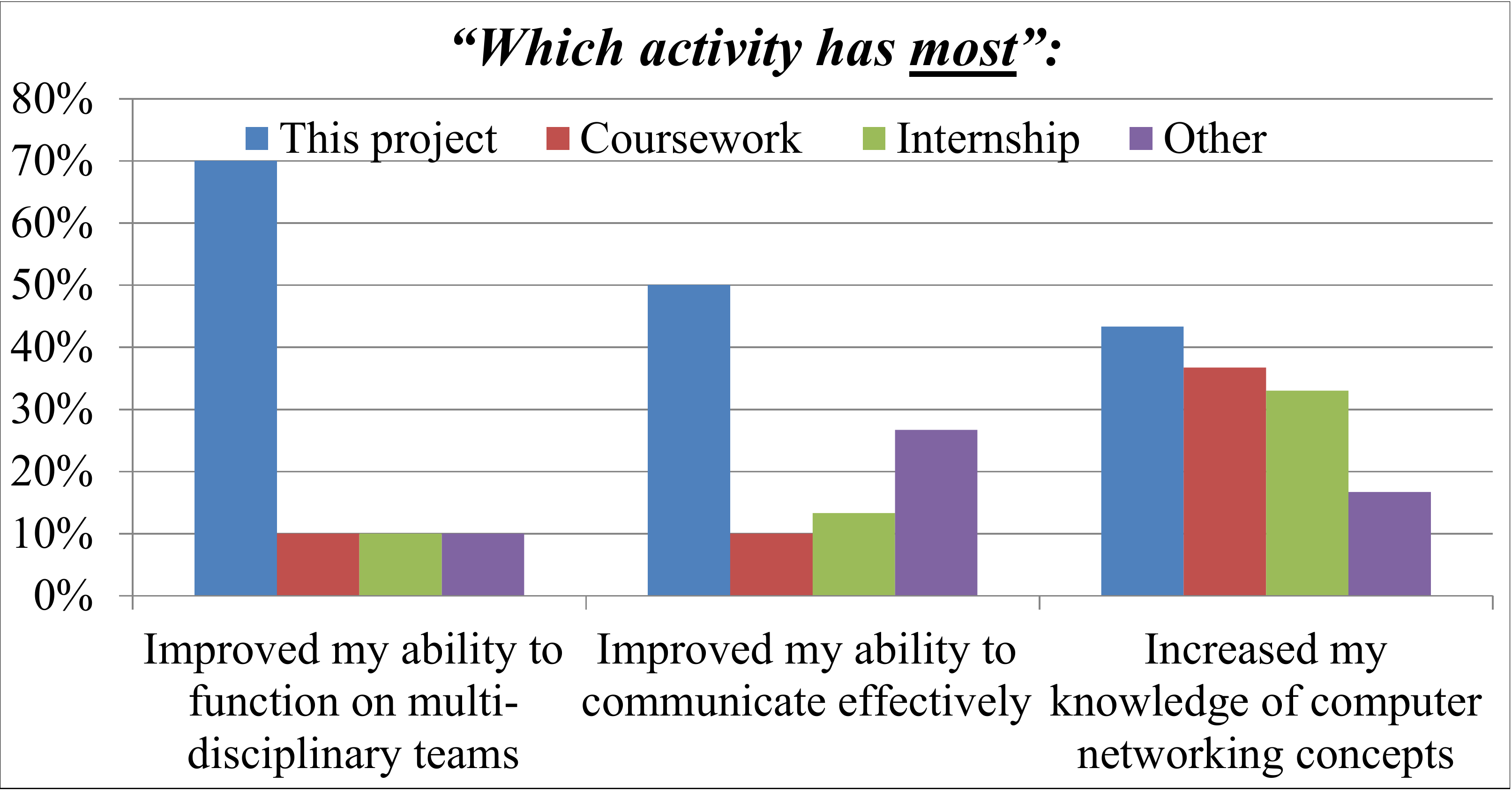}
\caption{\label{fig:ComparedToOtherActivities} Project survey results: Project involvement compared with other activities.}
\end{figure}

\begin{figure}
\centering
\includegraphics[width=\columnwidth]{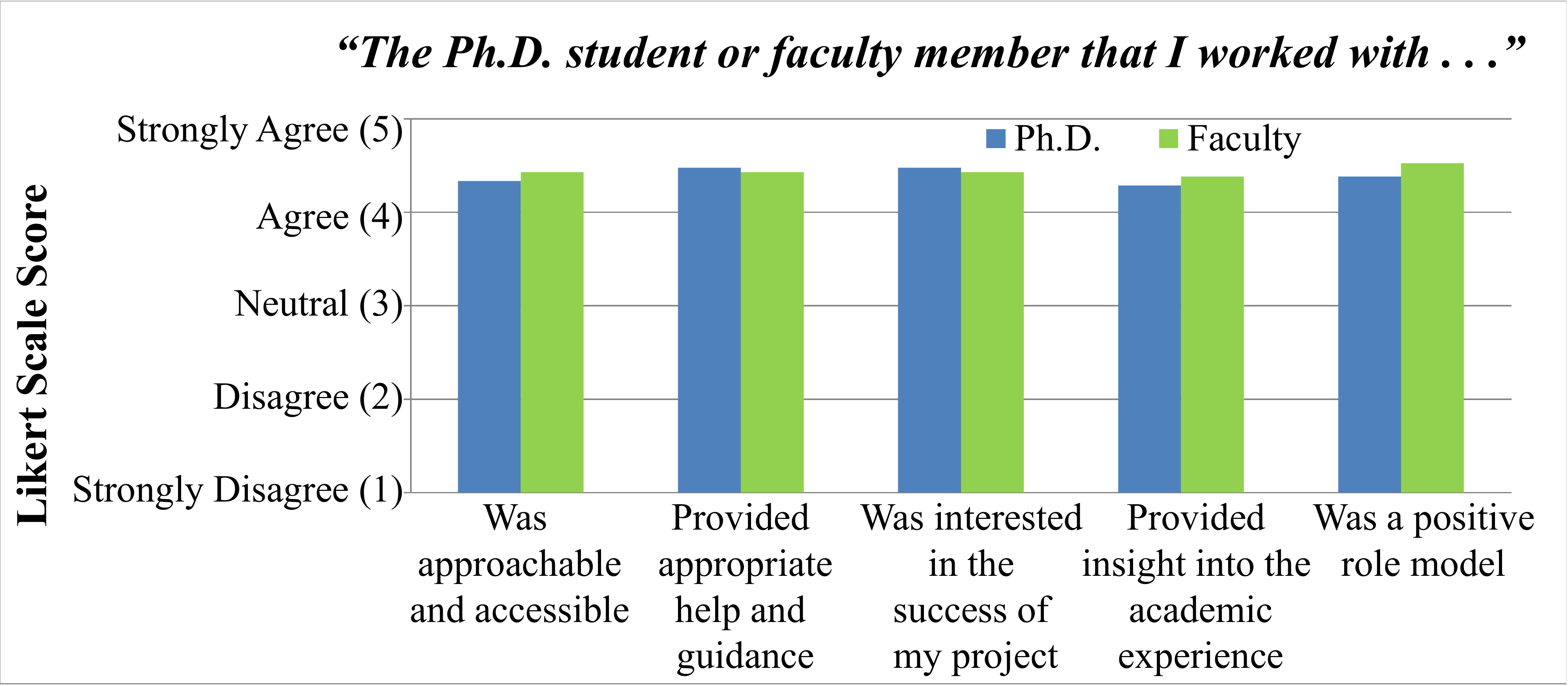}
\caption{\label{fig:compareProfStudent} Project survey results: Evaluation of mentoring by Ph.D.\ students and faculty.}
\end{figure}

The survey response rate was 75.5\%.
In the survey's open-ended questions, students shared many observations, comments, and suggestions about the project organization. Fig.~\ref{fig:SurveyHighlights}, \ref{fig:ComparedToOtherActivities}, and \ref{fig:compareProfStudent} show some of the results.
\makered{Fig.~\ref{fig:SurveyHighlights} demonstrates the percentage of students that agreed (or strongly agreed) and disagreed (or strongly disagreed) with statements related to the educational objectives (see Section~\ref{sect:Objectives}) and overall student experience. Fig.~\ref{fig:ComparedToOtherActivities} shows how the students compared their project involvement to other activities (e.g., coursework, internship)
on particular aspects of soft skills and technical knowledge. Fig.~\ref{fig:compareProfStudent} shows on the Likert scale the average evaluation score for some of the survey questions related to the student mentorship by faculty members and Ph.D.\ students.}


\newcommand \indsize {-2.5em}

\begin{table*}[t]
\center
\scriptsize
\begin{tabular}{|p{0.14\columnwidth}||p{0.68\columnwidth}|p{1.06\columnwidth}|}
\hline & Undergraduate & M.S.\ \\
\hline
\hline Academic
 &
\textbf{M.S.\ Students} --  Columbia (1), University of Michigan (1);  
\textbf{Ph.D.\ Candidates} --  Columbia (1), Princeton  \textbf{(2)}, Texas A\&M (1), UT Austin (1)
 &
\textbf{Ph.D.\ Candidates} --
 Carnegie Mellon (1), Columbia \textbf{(6)}, Harvard (1), Northwestern (1)
  \\
\hline Software Roles
 &
Program Manager at Microsoft; Manager of Cloud Operations at Bullhorn; Associate Applications Developer at AT\&T; Flight Software Engineer at Marshall Space Flight Center\newline
\textbf{Software Developers} -- Akcelita (1), Bloomberg LP (1), Credit~Karma~(1), ZocDoc \textbf{(2)}
 &
iOS Developer at Longtail Studios; Cofounder of Software Company (Vessel.io); Technology Associate at JPMorgan Chase; Applications Engineer at OPNET; Sales Engineer at Bluebox Security\newline
\textbf{Software Developers} --  Apple (1), Interactive Brokers (1), Lam Research (1), Microsoft (1), Oracle (1), VMWare (1), VM Turbo (1), Work Market (1)
  \\
\hline Hardware Roles &
Battery Test Engineer at 24M Technologies;
Product Engineer at Intel;
Electrical Engineer at Bala Consulting Engineers
 &
Systems Engineers at Raytheon Co. \textbf{(2)}; Electrical Engineer at SunPower Corp.\newline
\textbf{Circuit Design Engineers} -- Ambarella (1), SK Hynix (1), Credit Suisse (1), Silicon Laboratories (1), Boeing Space Systems (1) 
 \\
\hline Other &
Unknown (3)
 &
Pilot at U.S. Air Force; 
Business Assistant at TCL Multimedia;
Unknown (1)
 \\
\hline
\end{tabular}
\caption{\label{tab:plac}The placements of 22 undergraduate and 32 M.S.\ students that already graduated (6 undergraduate and 5 M.S.\ students are are yet to graduate).}
\end{table*}

Overall, the students' experiences were overwhelmingly positive. Over 90\% of the students
believed their project experience to be rewarding and enriching. Over 85\% of the students indicated that working on this project improved their ability to function on multidisciplinary teams and to communicate effectively, made them a better computer scientist or a better engineer, and was a valuable part of their education (Fig.~\ref{fig:SurveyHighlights}).
70\% of the students indicated that working on the project improved their ability to function on multidisciplinary teams \emph{more than any other activity}. Additionally, 50\% of the students indicated that working on the project improved their ability to communicate effectively more than any other activity, and over 40\% of the students indicated that this project increased their knowledge of computer networking more than any other activity (Fig.~\ref{fig:ComparedToOtherActivities}).

The students' feedback on the project organization features provided additional insights into the features' effectiveness.
\begin{myitemize}
\item  \textbf{Multidisciplinary projects:} Most students enjoyed the multidisciplinary nature of their projects. When specifying what they liked most about the project, over~50\% of the students commented on one of its multidisciplinary aspects. One student enjoyed her project being \emph{``about both hardware and software"}, and said it was \emph{``innovative and challenging to integrate many different aspects in one"}. One student's favorite thing about his project was the \emph{``integration of my work with other parts of the system -- felt like a cohesive project that mattered more"}. 
    However, students also noted that \emph{``getting everyone on same page was difficult at times"}, and said that \emph{``not being able to know exactly what others are doing''} was an impediment to achieving some of their project goals. Based on this feedback, we have been encouraging the students to independently and proactively collaborate with each other.

\item  \textbf{Real-world system integration deadlines:}
    \makered{While students nearly universally enjoyed being able to integrate their work with the work of others, students' opinions about the deadlines were mixed.
    Some of the students said that they liked the \emph{"feel of product evolution''} and \emph{"liked the challenging deadlines''}. 
    Yet other students pointed out several deadline-related inefficiencies in project organization. For example, one of the students noted that at times he was not given enough advance notice on system requirements prior to a deadline. Another student mentioned that the deadlines resulted in unnecessarily ``long and boring'' system testing. Another student also felt that because of too many integration deadlines she did not have enough time to make progress on her components' next-generation design.
    To improve the student's experiences, we dedicated extra attention to making sure that our subsequent deadlines (such as the Phase VI deadline in April 2013)
    were organized such that they did not negatively impact the students.} 

\item  \textbf{Frequent cross-group meetings:} Most students appreciated the opportunities for problem-solving and work discussions and showcasing provided by the regular cross-group meetings. One student noted that \emph{``the meetings are an excellent way for putting everything in the big picture."} 
    Yet several students also commented that the meetings were unnecessarily long, and suggested that a better meeting structure should be considered for future projects. To improve the quality of the meeting presentations we encourage students to discuss their presentation with their Ph.D.\ mentors. We are also considering introducing joint presentations for students from the same research group.

\item  \textbf{Frequent system demonstrations:}
    \emph{Over~95\% of the students} indicated that presenting their work was a rewarding and enriching experience. Several students specifically mentioned the presentation skills among the skills they acquired or improved while working on the umbrella project.
    One student noted that
    \emph{``the opportunity to present to
    others was invaluable. Plus it was a lot of fun!"}

\item  \textbf{Student mentorship:} The majority of students appreciated the
    support provided by their Ph.D.\ student mentors.
    Over~80\% of the students said that their mentor was approachable and accessible, and provided appropriate 
    guidance (Fig.~\ref{fig:compareProfStudent}).
    One student noted: \emph{``I've never considered pursuing higher education past undergrad level, but I'm starting to see the appeal in getting a Masters and/or Ph.D.\ now, after working around so many happy and passionate people."} Several students additionally specifically complimented their mentors.


\item  \textbf{Outreach and promotion:}
    \makered{Students were enthusiastic and excited about many different aspects of the
    opportunities to present their work. More than 80\% of the students said that this experience improved their communication skills and more than 50\% said that it helped them to clarify their career path. One of the students noted that presenting her work was an \emph{"overall confidence booster"}. Another specifically mentioned that it was \emph{``incredibly rewarding to start the project not knowing very much about the subject and end with the ability to present my work with real knowledge of what I had worked on"}}.

\item  \textbf{Knowledge transfer:} Despite students' favorable opinions of their mentors and despite their appreciation of the opportunities to teach others, 
 a large portion of the negative feedback focused on insufficient knowledge transfer. 
Several students commented on the insufficient technical introduction to their project. One student stated that \emph{``in the beginning I felt I didn't have enough support to ask very basic things"}, and another student noted that \emph{``a lot of work goes to waste if you are unable to successfully pass it on to the next person''}. Based on this feedback we have increased our efforts to ensure that the students create high-quality up-to-date documentation.
\end{myitemize}

\makered{Following the feedback from the survey, as we worked through the Phase~VI demonstration of the umbrella project and past it, we have been modifying the organizational aspects of the project to additionally facilitate student development. 
We take added care in scheduling tasks related to system integration deadlines, so as not to over-burden students and detract from research and educational objectives.
To address cross-group technological issues more quickly and to improve the effectiveness of weekly cross-group meetings, 
we also conduct many more frequent (up to 2--3 times per week) smaller-group interactions, typically involving only the relevant students and their Ph.D.\ mentors. Additionally, we take added care to ease students' introduction to the project and to ensure other aspects of knowledge transfer. For example, we now educate students on important aspects of umbrella project disciplines that they may be unfamiliar with (e.g., we ensure that computer science students are trained to handle experimental electronics).}

\section{Outcomes}\label{sect:outcomes}
In this section, we list the outcomes of the project including project awards, individual awards, student publications, and career placements.

\noindent\textbf{Project Awards:} A demonstration of the
EnHANTs testbed (with 10 student authors and 4 faculty mentors) received
the Best Student Demo Award at the ACM SenSys'11 conference
\cite{SenSys2011Demo}. The EnHANTs project also won the 1$^\textrm{st}$
place in the 2009 Vodafone Americas Foundation Wireless Innovation
Competition and a project vision paper (with 3 students acknowledged)
received the 2011 IEEE Communications Society Award for Advances in
Communications \cite{Gorlatova_Enhants_wircom}. Following these
recognitions, the EnHANTs project and the student outcomes were also
featured in several media outlets including BBC/PRI, RFID Journal,
Columbia's The Record, IEEE Spectrum, and MIT Technology Review.

\noindent\textbf{Individual Awards:} Students received: 
NSF Graduate Research Fellowship, 2013 MIT EECS Rising Star workshop
invitation, Columbia Presidential Teaching Award, Wei Family Foundation
Fellowship (stipend and tuition for an incoming Ph.D.\ student),
Exzellenzstipendium 2011 Fellowship (competitive Austrian fellowship),
as well as departmental Undergraduate Research Awards (two), and M.S.\ Awards of
Excellence (two). The Ph.D.\ mentors received several awards, some 
directly stemming from mentoring activities including the Google
Anita Borg Fellowship, NSF Graduate Research Fellowship, Computing
Innovation Fellowship as well as departmental Jury Award, Armstrong
Award, and Collaborative Research Award. 

\noindent\textbf{Publications:} Nine students co-authored research
papers \cite{KymEncapsulation, kinetic_SIG, sarik_more_2013,
  afsar_evaluating_2012, Gorlatova2013Prototyping,
  Gorlatova_Infocom2011, Gorlatova_TMC13, Wang_TMC12}. A
undergraduate authored \cite{kinetic_SIG} (with four other students
acknowledged). This paper was featured in the MIT Technology Review Report on
Business \cite{internetOfYouMITTechn} and in the MIT Technology Review
Physics ArXiv Blog \cite{humanMotionMITTechn}.  29 students co-authored
published demo descriptions
\cite{SeconDemoEnHANTs2010,MobiComDemo,MobiSys2011Demo,SenSys2011Demo,IDTechDemo2012,Margolies2013Demo}
(see Section \ref{sect:Umbrella}) and five students co-authored 
first-of-their-kind energy-harvesting datasets
\cite{columbia-enhants-light-energy-traces,columbia-kinetic-2014-05-13}.

\noindent\textbf{Career Placement:}
As of September 2014, 83\% of the B.S.\ and M.S. students have graduated and the other 17\% are continuing their
studies. Table~\ref{tab:plac} shows the current positions of the
graduated students.
Of the students who have already graduated, 
30\% continued to
higher-level academic programs in top schools like Princeton, Harvard,
and Carnegie Mellon, \emph{including seven at Columbia (some of which
  later became mentors in the program)}.\footnote{All of the M.S.\ students were originally enrolled in an  M.S.\ only
  program that does not lead to a Ph.D.\ and \emph{the high percentage
    of Ph.D.\ acceptance among them is very unique}.} 
The other 70\% were recruited by technology
companies, such as Microsoft, Raytheon, Boeing, Apple, Intel, OPNET,
Oracle, VMturbo, and ZocDoc, as well as by NASA and the US Air
Force. Several students have indicated that working on the umbrella project prepared them for some of the challenges they face in their careers. For example, one student noted that the \emph{``experience presenting my work has been really helpful in my current job profile"}, and another highlighted that \emph{``being held accountable for deadlines and project completeness helped prepare me for work environment"}.

In addition, a few of the Ph.D.\ mentors already graduated and assumed
postdoctoral positions in EPFL, Princeton, and Columbia, research
positions in AT\&T Labs Research and D.E.\ Shaw Research, and industry
positions. Three of the mentors also founded start-ups. All of the
mentors indicated that their mentoring experience was instrumental in
shaping their career paths.

\section{Conclusions}
\label{sect:Conclusion}

%

While the modern computing landscape increasingly requires large-scale system engineering skills, such skills are rarely acquired in a typical computing program. To address this, over the last 5 years, we have been engaging a diverse group of students in research projects within a large-scale interdisciplinary \emph{Energy Harvesting Active Networked Tags (EnHANTs)} project.
To date, 180 \mbox{semester-long} projects have been completed within the EnHANTs project. The projects challenge students' knowledge and organizational and communication skills. Some of the approaches we have used to facilitate student learning are the \mbox{real-world} system development constraints, regular cross-group meetings, and extensive involvement of Ph.D.~students in student mentorship and knowledge transfer. Students find the projects rewarding and gain valuable skills. Our experience demonstrates feasibility of engaging diverse groups of students on large-scale interdisciplinary research efforts. It sheds light on some potential pitfalls of such efforts (e.g., inadequate cross-group communication and knowledge transfer), and suggests best practices to overcome these challenges.

\section*{Acknowledgments}
The authors thank the students for their participation in the project and the evaluation survey as well as all of the Ph.D.\ students who participated in the project through their mentorship of the student projects.

\bibliographystyle{IEEEtran}
\bibliography{maria,sarik,rob}  

\begin{thebibliography}{10}
\providecommand{\url}[1]{#1}
\csname url@samestyle\endcsname
\providecommand{\newblock}{\relax}
\providecommand{\bibinfo}[2]{#2}
\providecommand{\BIBentrySTDinterwordspacing}{\spaceskip=0pt\relax}
\providecommand{\BIBentryALTinterwordstretchfactor}{4}
\providecommand{\BIBentryALTinterwordspacing}{\spaceskip=\fontdimen2\font plus
\BIBentryALTinterwordstretchfactor\fontdimen3\font minus
  \fontdimen4\font\relax}
\providecommand{\BIBforeignlanguage}[2]{{%
\expandafter\ifx\csname l@#1\endcsname\relax
\typeout{** WARNING: IEEEtran.bst: No hyphenation pattern has been}%
\typeout{** loaded for the language `#1'. Using the pattern for}%
\typeout{** the default language instead.}%
\else
\language=\csname l@#1\endcsname
\fi
#2}}
\providecommand{\BIBdecl}{\relax}
\BIBdecl

\bibitem{schocken2012taming}
S.~Schocken, ``Taming complexity in large-scale system projects,'' in
  \emph{Proc. ACM SIGCSE'12}, Feb. 2012.

\bibitem{wolf2000embedded}
W.~Wolf and J.~Madsen, ``Embedded systems education for the future,''
  \emph{Proc. of the IEEE}, vol.~88, no.~1, pp. 23--30, 2000.

\bibitem{lee2010project}
C.~S. Lee, J.~H. Su, K.~E. Lin, J.~H. Chang, and G.~H. Lin, ``A project-based
  laboratory for learning embedded system design with industry support,''
  \emph{IEEE Trans. Educ.}, vol.~53, no.~2, pp. 173--181, 2010.

\bibitem{hussmann2007crazy}
S.~Hussmann and D.~Jensen, ``Crazy car race contest: Multicourse design
  curricula in embedded system design,'' \emph{IEEE Trans. Educ.}, vol.~50,
  no.~1, pp. 61--67, 2007.

\bibitem{Bernat2000structuring}
A.~Bernat, P.~J. Teller, A.~Gates, and N.~Delgado, ``Structuring the student
  research experience,'' in \emph{Proc. ACM ITiCSE'00}, July 2000.

\bibitem{wenderholm2004challenges}
E.~Wenderholm, ``Challenges and the elements of success in undergraduate
  research,'' \emph{ACM SIGCSE Bull.}, vol.~36, no.~4, pp. 73--75, 2004.

\bibitem{Coleman2012Collaboration}
B.~Coleman and M.~Lang, ``Collaboration across the curriculum: A disciplined
  approach to developing team skills,'' in \emph{Proc. ACM SIGCSE'12}, Feb.
  2012.

\bibitem{raicu2009enhancing}
D.~S. Raicu and J.~D. Furst, ``Enhancing undergraduate education: A {REU} model
  for interdisciplinary research,'' \emph{ACM SIGCSE Bull.}, vol.~41, no.~1,
  pp. 468--472, 2009.

\bibitem{Gorlatova_Enhants_wircom}
M.~Gorlatova, P.~Kinget, I.~Kymissis, D.~Rubenstein, X.~Wang, and G.~Zussman,
  ``Energy-harvesting active networked tags ({EnHANTs}) for ubiquitous object
  networking,'' \emph{IEEE Wireless Commun.}, vol.~17, no.~6, pp. 18--25, Dec.
  2010.

\bibitem{EnHANTsProject}
``Energy harvesting active networked tags ({EnHANTs}) project,''
  \url{http://enhants.ee.columbia.edu}, 2014.

\bibitem{SeconDemoEnHANTs2010}
M.~Gorlatova, T.~Sharma, D.~Shrestha, E.~Xu, J.~Chen, A.~Skolnik, D.~Piao,
  P.~Kinget, I.~Kymissis, D.~Rubenstein, and G.~Zussman, ``{Demo: Prototyping
  Energy Harvesting Active Networked Tags (EnHANTs) with MICA2 Motes},'' in
  \emph{Proc. IEEE SECON'10}, June 2010.

\bibitem{MobiComDemo}
M.~Gorlatova, Z.~Noorbhaiwala, A.~Skolnik, J.~Sarik, M.~Szczodrak, J.~Chen,
  M.~Zapas, L.~Carloni, P.~Kinget, I.~Kymissis, D.~Rubenstein, and G.~Zussman,
  ``Demo: Prototyping energy harvesting active networked tags,'' \emph{Demo
  presented in ACM MobiCom'10}, Sept. 2010.

\bibitem{SenSys2011Demo}
G.~Stanje, P.~Miller, J.~Zhu, A.~Smith, O.~Winn, R.~Margolies, M.~Gorlatova,
  J.~Sarik, M.~Szczodrak, B.~Vigraham, L.~Carloni, P.~Kinget, I.~Kymissis, and
  G.~Zussman, ``Demo: Organic solar cell-equipped energy harvesting active
  networked tag ({EnHANT}) prototypes,'' in \emph{Proc. ACM SenSys'11}, Nov.
  2011, {B}est {S}tudent {D}emo {A}ward.

\bibitem{MobiSys2011Demo}
J.~Zhu, G.~Stanje, R.~Margolies, M.~Gorlatova, J.~Sarik, Z.~Noorbhaiwala,
  P.~Miller, M.~Szczodrak, B.~Vigraham, L.~Carloni, P.~Kinget, I.~Kymissis, and
  G.~Zussman, ``Demo: Prototyping {UWB}-enabled {EnHANTs},'' in \emph{Proc. ACM
  MobiSys'11}, June 2011.

\bibitem{IDTechDemo2012}
M.~Wang, L.~Pena, J.~Sarik, H.~Wang, R.~Margolies, M.~Gorlatova, G.~Burrow,
  B.~Vigraham, P.~Kinget, I.~Kymissis, and G.~Zussman, ``Energy harvesting
  active networked tags ({EnHANTs}) prototypes,'' Nov. 2012, {D}emo presented
  in IDtechEx Energy Harvesting and Storage, {Washington, DC}.

\bibitem{Margolies2013Demo}
R.~Margolies, L.~Pena, K.~Kim, Y.~Kim, M.~Wang, M.~Gorlatova, J.~Sarik, J.~Zhu,
  P.~Kinget, I.~Kymissis, and G.~Zussman, ``Demo: An adaptive testbed of energy
  harvesting active networked tags ({EnHANTs}) prototypes,'' in \emph{Proc.
  IEEE INFOCOM'13}, Apr. 2013.

\bibitem{Gorlatova2013Prototyping}
M.~Gorlatova, R.~Margolies, J.~Sarik, G.~Stanje, J.~Zhu, B.~Vigraham,
  M.~Szczodrak, L.~Carloni, P.~Kinget, I.~Kymissis, and G.~Zussman,
  ``Prototyping energy harvesting active networked tags ({EnHANTs}),'' in
  \emph{Proc. IEEE INFOCOM'13 mini-conference}, Apr. 2013.

\bibitem{Gorlatova_EnHANTS_TOSN}
------, ``Energy harvesting active networked tags ({EnHANTs}): Prototyping and
  experimentation,'' \emph{Submitted to ACM Transactions on Sensor Networks
  (TOSN)}, 2014.

\bibitem{polack2006learning}
J.~A. Polack-Wahl and K.~Anewalt, ``Learning strategies and undergraduate
  research,'' \emph{ACM SIGCSE Bull.}, vol.~38, no.~1, pp. 209--213, 2006.

\bibitem{Greening2002Undergraduate}
T.~Greening and J.~Kay, ``Undergraduate research experience in computer science
  education,'' in \emph{Proc. ACM ITiCSE'02}, June 2002.

\bibitem{peckham2007increasing}
J.~Peckham, P.~Stephenson, J.~Herv{\'e}, R.~Hutt, and M.~Encarna{\c{c}}{\~a}o,
  ``Increasing student retention in computer science through research programs
  for undergraduates,'' \emph{ACM SIGCSE Bull.}, vol.~39, no.~1, pp. 124--128,
  2007.

\bibitem{saliklis2006putting}
E.~P. Saliklis, ````putting a fence'' around architectural engineering
  undergraduate research projects,'' in \emph{ASEE National Conference}, June
  2006.

\bibitem{savage2007integrating}
R.~N. Savage, K.~C. Chen, and L.~Vanasupa, ``Integrating project-based learning
  throughout the undergraduate engineering curriculum,'' \emph{Journal of
  {STEM} Education: Innovations \& Research}, vol.~8, no. 3/4, p.~15, July-Dec.
  2007.

\bibitem{dym2005engineering}
C.~L. Dym, A.~M. Agogino, O.~Eris, D.~D. Frey, and L.~J. Leifer, ``Engineering
  design thinking, teaching, and learning,'' \emph{Journal of Engineering
  Education}, vol.~94, no.~1, pp. 103--120, 2005.

\bibitem{hadim2002enhancing}
H.~A. Hadim and S.~K. Esche, ``Enhancing the engineering curriculum through
  project-based learning,'' in \emph{Proc. IEEE Frontiers in Education}, 2002.

\bibitem{felder2000future}
R.~M. Felder, D.~R. Woods, J.~E. Stice, and A.~Rugarcia, ``The future of
  engineering education ii. teaching methods that work,'' \emph{Chemical
  Engineering Education}, vol.~34, no.~1, pp. 26--39, 2000.

\bibitem{mills2003engineering}
J.~E. Mills and D.~F. Treagust, ``Engineering education: Is problem-based or
  project-based learning the answer?'' \emph{Australasian Journal of
  Engineering Education}, vol.~3, pp. 2--16, 2003.

\bibitem{zydney2002impact}
A.~L. Zydney, J.~S. Bennett, A.~Shahid, and K.~W. Bauer, ``Impact of
  undergraduate research experience in engineering,'' \emph{Journal of
  Engineering Education}, vol.~91, no.~2, pp. 151--157, 2002.

\bibitem{ABET2000}
ABET, ``{Engineering Change}: A study of the impact of {EC2000},''
  \url{www.abet.org/engineering-change}, 2006.

\bibitem{Crepaldi2011}
M.~Crepaldi, C.~Li, K.~Dronson, J.~Fernandes, and P.~Kinget, ``An interference
  robust {S-OOK IR-UWB} transceiver chipset,'' \emph{IEEE J. Solid-State
  Circuits}, vol.~46, no.~10, pp. 2284--2299, Oct. 2011.

\bibitem{MicaMote}
``Crossbow {MICA2} {M}ote,''
  \url{www.xbow.com/Products/Product_pdf_files/Wireless_pdf/MICA2_Datasheet.pdf},
  2009.

\bibitem{HarlemChildrenSociety}
``Harlem children society,'' \url{harlemchildrensociety.org}, 2014.

\bibitem{KymEncapsulation}
M.~E. Bhalke, S.~P. Subbarao, and I.~Kymissis, ``A laboratory process for
  encapsulation of air sensitive organic devices,'' \emph{IEEE Trans. Electron
  Devices}, vol.~57, no.~1, pp. 153--156, Jan. 2010.

\bibitem{kinetic_SIG}
M.~Gorlatova, J.~Sarik, G.~Grebla, M.~Cong, I.~Kymissis, and G.~Zussman,
  ``Movers and shakers: Kinetic energy harvesting for the internet of things,''
  in \emph{Proc. ACM SIGMETRICS'14}, June 2014.

\bibitem{sarik_more_2013}
J.~Sarik, K.~Kim, M.~Gorlatova, I.~Kymissis, and G.~Zussman, ``More than meets
  the eye - a portable measurement unit for characterizing light energy
  availability,'' in \emph{Proc. {IEEE} {GlobalSIP'13} Symp. on Energy
  Harvesting and Green Wireless Communications}, Dec. 2013.

\bibitem{afsar_evaluating_2012}
Y.~Afsar, J.~Sarik, M.~Gorlatova, G.~Zussman, and I.~Kymissis, ``Evaluating
  photovoltaic performance indoors,'' in \emph{Proc. 38th {IEEE} Photovoltaic
  Specialist Conference ({PVSC38)}}, June 2012.

\bibitem{Gorlatova_Infocom2011}
M.~Gorlatova, A.~Wallwater, and G.~Zussman, ``Networking low-power energy
  harvesting devices: Measurements and algorithms,'' in \emph{Proc. IEEE
  INFOCOM'11}, Apr. 2011.

\bibitem{Gorlatova_TMC13}
------, ``Networking low-power energy harvesting devices: Measurements and
  algorithms,'' \emph{IEEE Trans. Mobile Comput.}, vol.~12, no.~9, pp.
  1853--1865, Sept. 2013.

\bibitem{Wang_TMC12}
Z.~Wang, A.~Tajer, and X.~Wang, ``Communication of energy harvesting tags,''
  \emph{IEEE Trans. Commun.}, vol.~60, no.~4, pp. 1159--1166, Apr. 2012.

\bibitem{internetOfYouMITTechn}
\BIBentryALTinterwordspacing
R.~Metz, ``The {I}nternet of you,'' \emph{MIT Technology Review}, May 2014.
  [Online]. Available:
  \url{http://www.technologyreview.com/news/527386/the-internet-of-you/}
\BIBentrySTDinterwordspacing

\bibitem{humanMotionMITTechn}
\BIBentryALTinterwordspacing
{The Physics Arxiv Blog}, ``Human motion will power the {I}nternet of things,
  say energy harvesting engineers,'' \emph{MIT Technology Review}, Jul. 2013.
  [Online]. Available:
  \url{http://m.technologyreview.com/view/516816/human-motion-will-power-the-internet-of-things-say-energy-harvesting-engineers/}
\BIBentrySTDinterwordspacing

\bibitem{columbia-enhants-light-energy-traces}
M.~Gorlatova, M.~Zapas, E.~Xu, M.~Bahlke, I.~J. Kymissis, and G.~Zussman,
  ``{CRAWDAD} data set columbia/enhants (v. 2011-04-07),''
  http://crawdad.cs.dartmouth.edu/columbia/enhants, Apr. 2011.

\bibitem{columbia-kinetic-2014-05-13}
M.~Cong, K.~Kim, M.~Gorlatova, J.~Sarik, J.~Kymissis, and G.~Zussman,
  ``{CRAWDAD} data set columbia/kinetic (v. 2014-05-13),''
  http://crawdad.org/columbia/kinetic/, May 2014.

\end{thebibliography}
%
%

\end{document}